\documentclass[aps,pra,showpacs,amsmath,amssymb,amsfonts,tightenlines,twocolumn,lengthcheck]{revtex4-1}
\usepackage{amsmath}
\usepackage{amsfonts}
\usepackage{graphicx}
\usepackage{epsfig}
\usepackage{color}
\usepackage{txfonts}
\usepackage[colorlinks,citecolor=blue]{hyperref}

\begin{document}

\title{All-optical quantum simulation of ultrastrong optomechanics}
\author{Xian-Li Yin}
\affiliation{Key Laboratory of Low-Dimensional Quantum Structures and Quantum Control of Ministry of Education, Key Laboratory for Matter Microstructure and Function of Hunan Province, Department of Physics and Synergetic Innovation Center for Quantum Effects and Applications, Hunan Normal University, Changsha 410081, China}
\author{Yue-Hui Zhou}
\affiliation{Key Laboratory of Low-Dimensional Quantum Structures and Quantum Control of Ministry of Education, Key Laboratory for Matter Microstructure and Function of Hunan Province, Department of Physics and Synergetic Innovation Center for Quantum Effects and Applications, Hunan Normal University, Changsha 410081, China}
\author{Jie-Qiao Liao}
\email{Corresponding author: jqliao@hunnu.edu.cn}
\affiliation{Key Laboratory of Low-Dimensional Quantum Structures and Quantum Control of Ministry of Education, Key Laboratory for Matter Microstructure and Function of Hunan Province, Department of Physics and Synergetic Innovation Center for Quantum Effects and Applications, Hunan Normal University, Changsha 410081, China}

\begin{abstract}
The observation of single-photon optomechanical effects is a desired task in cavity optomechanics. However, the realization of ultrastrong optomechanical interaction remains a big challenge. Here, we present an all-optical scheme to simulate ultrastrong optomechanical coupling based on a Fredkin-type interaction, which consists of two exchange-coupled modes with the coupling strength depending on the photon number in another controller mode. This coupling enhancement is assisted by the displacement amplification according to the physical idea of the Bogoliubov approximation, which is realized by utilizing a strong driving to pump one of the two exchanging modes. Our numerical simulations demonstrate that the enhanced optomechanical coupling can enter the single-photon strong-coupling and even ultrastrong-coupling regimes. We also show the creation of macroscopic quantum superposed states and the implementation of a weak-to-strong transition for quantum measurement in this system. This work will pave the way to quantum simulation of single-photon optomechanical effects with current experimental platforms.
\end{abstract}

\date{\today}
\maketitle
\section{Introduction}
The recent developments in cavity optomechanics~\cite{Kippenberg2008rev,Aspelmeyer2012rev,Aspelmeyer2014} facilitate the study of radiation-pressure interactions between electromagnetic fields and mechanical vibrations, especially at the single-photon level~\cite{Rabl2011,Nunnenkamp2011,Liao2012,Liao2013,Liao2013PRA,Hong2013,Xu2013,Tang2014,Marshall2003,Liao2016}. The coherent nonlinear optomechanical interaction is the physical origin of various interesting physical effects, such as the appearance of phonon sidebands in the cavity emission spectrum~\cite{Nunnenkamp2011,Liao2012}, photon blockade effect induced by the moving boundary~\cite{Rabl2011,Liao2013,Liao2013PRA,Xu2013}, and the generation of macroscopic quantum superposition~\cite{Marshall2003,Liao2016}. In particular, this provides a good platform for investigating some fundamental issues in quantum mechanics, for example, the quantum measurement problem, a puzzle that has not yet been completely resolved~\cite{Neumann2018}. Hence, how to efficiently manipulate the optomechanical interactions at the single-photon level has become one of the most interesting research topics in this field. However, it has remained a big challenge to observe single-photon optomechanical effects with current experimental techniques. This is because the magnitude of the optomechanical coupling associated with a single photon is too weak to enter the single-photon strong-coupling regime.

By far, several schemes have been proposed to amplify the single-photon optomechanical effects~\cite{Brennecke2008,Xuereb2012,Rimberg2014,Heikkila2014,Pirkkalainen2015,Liao2014,Liao2015,Lue2015,Lemonde2016,Li2016,Wang2017,Liao2020}, such as the enhancement of the optomechanical coupling with a collective density excitation of the Bose-Einstein condensate~\cite{Brennecke2008} or collective modes~\cite{Xuereb2012}, the resonant enhancement induced by either coupling~\cite{Liao2014,Liao2016} or cavity-frequency~\cite{Liao2015} modulation, the utilization of the strong nonlinearity in the Josephson junctions~\cite{Rimberg2014,Heikkila2014,Pirkkalainen2015}, the effective coupling enhancement induced by either the squeezing transformation~\cite{Lue2015,Lemonde2016,Li2016} or the displacement transformation~\cite{Liao2020}, and the utilizing of delayed quantum feedback~\cite{Wang2017}. Nevertheless, there are no reports on the demonstrations of single-photon optomechanical effects.  Inspired by the motivation of simulating the experimentally inaccessible physical effects with accessible physical system~\cite{Nori2014}, in this paper we propose an all-optical scheme to implement a quantum simulation of a tunable optomechanical interaction, which can enter single-photon strong-coupling and ultrastrong-coupling regimes~\cite{Hu2015}. In this way, we are able to generate distinct macroscopic superposed states with this enhanced optomechanical coupling. As an important feature of this system, the obtained optomechanical interaction can be controlled on demand by choosing proper drivings. Therefore, by adjusting the interaction strength from weak to strong, this system is used to exhibit the quantum measurement transition from the weak value to the expectation value~\cite{Aharonov1998,Pan2020}.

The rest of this paper is organized as follows. In Sec.~\ref{Physical model} we introduce the physical model and present the Hamiltonians. In Sec.~\ref{Generalized H} we derive the approximate optomechanical Hamiltonian and evaluate the validity of the approximate Hamiltonian. In Secs.~\ref{Cat-state generation},~\ref{Weak-to-strong measurement}, and~\ref{photon blockade effect} we study the generation of the Schr\"{o}dinger cat states in mode $b$, the implementation of the weak-to-strong transition of quantum measurement, and the photon blockade effect in mode $a$, respectively. We present a discussion on the experimental implementation of this scheme in Sec.~\ref{Discussions} and summarize this work in Sec.~\ref{Conclusions}.

\section{Physical model}\label{Physical model}
We consider an all-optical platform to simulate an ultrastrong optomechanical interaction based on a Fredkin-type interaction~\cite{Milburn1989,Patel2016,Gao2019Nature}, which consists of three optical modes described by the annihilation (creation) operators $a$ $(a^{\dagger})$, $b$ $(b^{\dagger})$, and $c$ $(c^{\dagger})$, with the corresponding resonance frequencies $\omega_{a}$, $\omega_{b}$, and $\omega_{c}$. Here the Fredkin interaction takes the form of a conditional two-mode exchange coupling (a beam-splitter-type coupling between modes $b$ and $c$ possessing a coupling strength depending on the photon numbers in mode $a$). Mode $c$ is driven by a strong monochromatic laser field with amplitude $\Omega_{c}$ and frequency $\omega_{L}$. In a rotating frame with respect to $H_{0}=\omega_{L}(b^{\dagger}b+c^{\dagger}c)$, the Hamiltonian of the system reads $(\hbar=1)$
\begin{eqnarray}
\label{LHmamiltonian}
H_{\textrm{sys}}&=&\omega_{a}a^{\dagger}a+\Delta_{b}b^{\dagger}b+\Delta_{c}c^{\dagger}c+ga^{\dagger}a(b^{\dagger}c+c^{\dagger}b)\notag\\
&&+\Omega_{c}c^{\dagger}+\Omega_{c}^{\ast}c,
\end{eqnarray}
where $\Delta_{b}=\omega_{b}-\omega_{L}$ and $\Delta_{c}=\omega_{c}-\omega_{L}$ are the detunings and $g$ is the coupling strength of the Fredkin interaction. In Eq.~(\ref{LHmamiltonian}) we neglect the counterrotating terms $\{|\Omega _{c}|\hspace{0.1cm}||c||/(\omega _{L}+\omega _{c}),|\Omega _{c}^{\ast}|\hspace{0.1cm}||c^{\dagger }||/(\omega _{L}+\omega _{c})\}\ll \{|\Omega_{c}|\hspace{0.1cm}||c^{\dagger }||/(\omega _{L}-\omega _{c}),|\Omega _{c}^{\ast}|\hspace{0.1cm}||c||/(\omega _{L}-\omega _{c})\}$, where $||c||$ and $||c^{\dagger}||$ denote the norm of the operators $c$ and $c^{\dagger}$, respectively.

To include the dissipation of the system, we assume that the three optical modes are coupled to three individual Markovian reservoirs. Then the evolution of the system can be described by the quantum master equation
\begin{equation}
\label{mastereq}
\dot{\rho}=i[\rho,H_{\text{sys}}]+\sum_{o=a,b,c}\{\kappa_{o}(\bar{n}_{o}+1)\mathcal{D}[o]\rho+\kappa_{o}\bar{n}_{o}\mathcal{D}[o^{\dagger}]\rho\},
\end{equation}
where $\mathcal{D}[o]\rho=o\rho o^{\dagger}-(o^{\dagger}o\rho+\rho o^{\dagger}o)/2$ is the standard Lindblad superoperator. The parameters $\kappa_{a}$, $\kappa_{b}$, $\kappa_{c}$ and $\bar{n}_{a}$, $\bar{n}_{b}$, $\bar{n}_{c}$ are, respectively, the damping rate and the environment thermal excitation occupation of modes $a$, $b$, $c$. Here, we consider the general case where the Markovian reservoirs are at finite temperatures. Though our scheme is based on three optical modes, it is universal to other bosonic systems, in which the effect of temperature may appear. Therefore, we include this effect in our simulation scheme.

\section{The generalized optomechanical Hamiltonian}\label{Generalized H}
In this section, we derive the generalized optomechanical Hamiltonian (approximate Hamiltonian) and evaluate the validity of the approximate Hamiltonian.

\subsection{Derivation of the approximate Hamiltonian}
We first sketch the inspiration concerning this coupling enhancement. Under the strong driving, the average photon number in mode $c$ is large and then the operator $c$ can be expressed as a sum of its average value and quantum fluctuation: $c\rightarrow c-\xi$. Consequently, the Fredkin interaction becomes $ga^{\dagger}a(b^{\dagger}c+c^{\dagger}b)-ga^{\dagger}a(\xi b^{\dagger}+\xi^{\ast}b)$, which can be reduced to an amplified optomechanical interaction (the target term) by discarding the first term under proper conditions. The rigorous derivation is performed by making the transformation
\begin{equation}
\rho^{\prime}=D_{c}(\xi)\rho D^{\dagger}_{c}(\xi),
\end{equation}
where $D_{c}(\xi)=\exp{(\xi c^{\dagger}-\xi^{\ast}c)}$ is the displacement operator, with the displacement amplitude $\xi\equiv|\xi|e^{i\theta_{c}}$. In the displacement representation, the quantum master equation of the system takes the form
\begin{equation}
\label{excMasEq}
\dot{\rho}^{\prime}=i[\rho^{\prime},H_{\text{dis}}]
+\sum_{o=a,b,c}(\kappa_{o}(\bar{n}_{o}+1)\mathcal{D}[o]\rho^{\prime}
+\kappa_{o}\bar{n}_{o}\mathcal{D}[o^{\dagger}]\rho^{\prime}),
\end{equation}
where we introduce the displaced Hamiltonian
\begin{eqnarray}
\label{excHamiltonian}
H_{\text{dis}} &=&\omega _{a}a^{\dagger }a+\Delta _{b}b^{\dagger }b+\Delta
_{c}c^{\dagger }c+ga^{\dagger }a(b^{\dagger }c+c^{\dagger }b)  \nonumber \\
&&-ga^{\dagger }a(b^{\dagger }\xi +b\xi ^{\ast }),
\end{eqnarray}
with the transient displacement amplitude determined by $\dot{\xi}=-(i\Delta_{c}+\kappa_{c}/2)\xi +i\Omega_{c}$. The steady-state displacement amplitude is given by $\xi_{\text{ss}}=\Omega_{c}/(\Delta_{c}-i\kappa_{c}/2)$, which can be tuned by selecting driving parameters $\Omega_{c}$ and $\omega_{L}$. To investigate the ultrastrong optomechanics, hereafter we consider a few excitations in modes $a$ and $b$. Under the parameter conditions $\vert\Delta_{c}-\Delta_{b}\vert\gg gn_{a}\sqrt{n_{b}n_{c}}$ and $\Delta_{b}\sim g\vert\xi\vert n_{a}\sqrt{n_{b}}$,  with $n_{o=a,b,c}$ the maximal dominant excitation numbers involved in mode $o$, the term $ga^{\dagger}a(b^{\dagger}c+c^{\dagger}b)$ can be ignored by the rotating-wave approximation (RWA) and mode $c$ decouples from modes $a$ and $b$. Apart from the decoupling term $\Delta_{c}c^{\dagger}c$, a generalized optomechanical Hamiltonian is obtained as
\begin{equation}
\label{appHamiltonian}
H_{\text{app}}=\omega_{a}a^{\dagger}a+\Delta_{b}b^{\dagger}b-g_{0}a^{\dagger}a(b^{\dagger }e^{i\theta_{c}}+be^{-i\theta_{c}}),
\end{equation}
where the interaction term takes the form as the product of the photon-number operator of mode $a$ and the rotated quadrature operator of mode $b$ (playing the role of the mechanical mode in typical optomechanical systems). Here the tunable single-photon optomechanical-coupling strength $g_{0}=g|\xi_{\text{ss}}|$ can be largely enhanced by using proper driving, and the phase angle $\theta_{c}$ is tunable by choosing the proper driving phase in $\Omega_{c}$.

\subsection{Evaluation of the validity of the approximate Hamiltonian}
The validity of the approximate Hamiltonian $H_{\text{app}}$ can be evaluated by checking the fidelity between the exact and approximate states, which are governed by the full Hamiltonian $H_{\text{dis}}$ and the approximate Hamiltonian $H_{\text{app}}$, respectively. To this end, we need to calculate the analytical results of the approximate and exact states for the system at time $t$.

\subsubsection{The analytical state determined by the approximate Hamiltonian $H_{\mathrm{app}}$}
To obtain the analytical state of the system at time $t$, we need to diagonalize the approximate Hamiltonian $H_{\text{app}}$. This can be done by introducing the displacement operator $D_{b}(\tilde{\beta})=e^{\tilde{\beta}(b^{\dagger }-b)}$, with the photon-number dependent displacement amplitude
\begin{equation}
\tilde{\beta}=\tilde{\beta}(a^{\dagger }a)=\frac{g_{0}a^{\dagger }a}{\Delta
_{b}}=\sum_{m=0}^{\infty }\tilde{\beta}(m)|m\rangle _{aa}\langle m|,
\end{equation}
where $\tilde{\beta}(m)=g_{0}m/\Delta _{b}$ is the $m$-photon-dependent displacement amplitude. In addition, we consider the phase angle $\theta_{c}=0$. Then the approximate Hamiltonian $H_{\text{app}}$ can be diagonalized as
\begin{equation}
\tilde{H}_{\text{app}}=D_{b}^{\dagger }(\tilde{\beta}) H_{\text{app}}D_{b}(\tilde{\beta}) =\omega _{a}a^{\dagger }a+\Delta_{b}b^{\dagger }b+\Delta _{c}c^{\dagger }c-\frac{g_{0}^{2}}{\Delta _{b}}
a^{\dagger }aa^{\dagger }a.
\end{equation}
Therefore, the analytical approximate state of the system at time $t$ can be obtained as
\begin{eqnarray}
|\Psi _{\text{app}}(t)\rangle  &=&U_{\text{app}}(t)\left\vert m\right\rangle
_{a}\left\vert \beta _{0}\right\rangle _{b}\left\vert \eta _{0}\right\rangle
_{c}  \nonumber \\
&=&D_{b}(\tilde{\beta}) e^{-i\tilde{H}_{\text{app}
}t}D_{b}^{\dagger }( \tilde{\beta}) \left\vert m\right\rangle
_{a}\left\vert \beta _{0}\right\rangle _{b}\left\vert \eta _{0}\right\rangle
_{c}  \nonumber \\
&=&\exp [-i\Theta _{\text{app}}^{(m)}\left( t\right) ]\left\vert
m\right\rangle _{a}\left\vert \beta _{1}(m)\right\rangle _{b}\left\vert \eta
_{1}\right\rangle _{c},
\end{eqnarray}
where we introduce the displacement amplitudes
\begin{subequations}
\begin{align}
\beta_{1}(m)&=\beta_{0}e^{-i\Delta_{b}t}+\frac{mg_{0}}{\Delta_{b}}(1-e^{-i\Delta_{b}t}),\\
\eta_{1}&=\eta_{0}e^{-i\Delta_{c}t}
\end{align}
\end{subequations}
and the phase
\begin{equation}
\Theta _{\text{app}}^{(m)}(t)=\omega_{a}mt-\frac{m^{2}g_{0}^{2}}{\Delta _{b}^{2}}[ \Delta _{b}t-\sin(\Delta _{b}t)]
-\frac{mg_{0}}{\Delta_{b}}\text{Im}[(e^{i\Delta_{b}t}-1)\beta_{0}^{\ast}].
\end{equation}

\subsubsection{The analytical state determined by the exact Hamiltonian $H_{\mathrm{dis}}$}
The exact state of the system at time $t$ can also be obtained analytically by diagonalizing the exact Hamiltonian $H_{\text{dis}}$. To this end, we introduce three transformation operators $T=e^{\lambda(b^{\dagger}c-c^{\dagger}b)}$, $D_{b}(\beta)=e^{\beta(b^{\dagger}-b)}$, and $D_{c}(\eta)=e^{\eta(c^{\dagger}-c)}$ based on the photon-number dependent mixing angle
\begin{equation}
\label{lambda}
\lambda=\lambda(a^{\dagger}a)=\frac{1}{2}\arctan \left(\frac{2ga^{\dagger }a}{\Delta _{c}-\Delta _{b}}\right)=\sum_{m=0}^{\infty}\lambda(m)|m\rangle_{aa}\langle m|
\end{equation}
and displacement amplitudes
\begin{subequations}
\label{betaandeta}
\begin{align}
\beta  &=\beta (a^{\dagger }a)=\frac{g_{0}a^{\dagger }a\cos \lambda }{\chi
_{b}}=\sum_{m=0}^{\infty }\beta (m)|m\rangle _{aa}\langle m|,\\
\eta  &=\eta (a^{\dagger }a)=\frac{g_{0}a^{\dagger }a\sin \lambda }{\chi
_{c}}=\sum_{m=0}^{\infty }\eta (m)|m\rangle _{aa}\langle m|,
\end{align}
\end{subequations}
with
\begin{subequations}
\label{chi}
\begin{align}
\chi _{b} &=\chi _{b}(a^{\dagger }a)=\sum_{m=0}^{\infty }\chi
_{b}(m)|m\rangle _{aa}\langle m|,  \nonumber \\
\chi _{c} &=\chi _{c}(a^{\dagger }a)=\sum_{m=0}^{\infty }\chi
_{c}(m)|m\rangle _{aa}\langle m|.
\end{align}
\end{subequations}
Here these $m$-photon dependent variables $\lambda(m)$, $\beta(m)$, $\eta(m)$, $\chi_{b}(m)$, and $\chi _{c}(m)$ are defined by
\begin{subequations}
\label{lambdambetaandetamchim}
\begin{align}
\lambda(m)&=\frac{1}{2}\arctan\left(\frac{2mg}{\Delta_{c}-\Delta _{b}}\right),\\
\beta(m)&=\frac{mg_{0}\cos[\lambda(m)]}{\chi _{b}(m)}, \\
\eta(m)&=\frac{mg_{0}\sin[\lambda(m)]}{\chi _{c}(m)},\\
\chi_{b}(m)\!&=\!\Delta_{b}\cos^{2}[\lambda(m)]\!+\!\Delta_{c}\sin^{2}[\lambda(m)]\!-\!mg\sin[2\lambda(m)],\\
\chi_{c}(m)\!&=\!\Delta_{b}\sin^{2}[\lambda(m)]\!+\!\Delta_{c}\cos^{2}[\lambda(m)]\!+\!mg\sin[2\lambda(m)].
\end{align}
\end{subequations}
By performing these transformations, the Hamiltonian $H_{\text{dis}}$ can be diagonalized as
\begin{eqnarray}
\label{diagonalized H}
\tilde{H}_{\text{dis}}&=&D_{c}^{\dagger}(\eta)D_{b}^{\dagger}(\beta)T^{\dagger }H_{\text{dis}}TD_{b}(\beta )D_{c}(\eta )\nonumber\\
&=&\omega_{a}a^{\dagger}a+\chi_{b}b^{\dagger}b+\chi_{c}c^{\dagger}c\nonumber \\
&&-\frac{g_{0}^{2}\cos^{2}\lambda}{\chi_{b}}a^{\dagger}aa^{\dagger}a-
\frac{g_{0}^{2}\sin^{2}\lambda}{\chi_{c}}a^{\dagger}aa^{\dagger}a.
\end{eqnarray}
Based on Eq.~(\ref{diagonalized H}), the exact state of the system at time $t$ corresponding to the initial state $\vert m\rangle_{a}\vert\beta_{0}\rangle_{b}\vert\eta _{0}\rangle_{c}$ can be obtained as
\begin{eqnarray}
\label{exact state at t}
|\Psi_{\text{ext}}(t)\rangle&=&U_{\text{ext}}(t)\vert m\rangle_{a}\vert\beta_{0}\rangle_{b}\vert\eta _{0}\rangle_{c}\nonumber \\
&=&TD_{b}(\beta)D_{c}(\eta)e^{-i\tilde{H}_{\text{dis}}t}D_{c}^{\dagger}(\eta)D_{b}^{\dagger}(\beta )T^{\dagger}\vert m\rangle_{a}\vert\beta_{0}\rangle_{b}\vert\eta_{0}\rangle_{c}\nonumber \\
&=&\exp [i\Theta^{(m)}_{\text{ext}}\left( t\right) ]\vert m\rangle_{a}\vert\beta_{2}(m)\rangle_{b}\vert\eta_{2}(m)\rangle_{c}.
\end{eqnarray}
Here, we introduce the displacement amplitudes
\begin{subequations}
\label{prime beta and prime eta}
\begin{align}
\beta _{2}(m) &=(\{\beta _{0}\cos [\lambda (m)]-\eta _{0}\sin [\lambda(m)]\}e^{-i\chi _{b}(m)t}  \nonumber \\
&+\beta (m)[1-e^{-i\chi _{b}(m)t}])\cos [\lambda (m)]  \nonumber \\
&+(\{\beta _{0}\sin [\lambda (m)]+\eta _{0}\cos [\lambda (m)]\}e^{-i\chi_{c}(m)t}  \nonumber\\
&+\eta (m)[1-e^{-i\chi _{c}(m)t}])\sin [\lambda (m)], \\
\eta _{2}(m) &=(\{\beta _{0}\sin [\lambda (m)]+\eta _{0}\cos [\lambda
(m)]\}e^{-i\chi _{c}(m)t}\nonumber \\
&+\eta (m)[1-e^{-i\chi _{c}(m)t}])\cos [\lambda (m)] \nonumber \\
&-(\{\beta _{0}\cos [\lambda (m)]-\eta _{0}\sin [\lambda (m)]\}e^{-i\chi_{b}(m)t}\nonumber \\
&+\beta (m)[1-e^{-i\chi _{b}(m)t}])\sin [\lambda (m)]
\end{align}
\end{subequations}
and the phase
\begin{eqnarray}
\label{ext phase angle}
\Theta_{\text{ext}}^{(m)}(t)&=&-m\omega_{a}t+\frac{g_{0}^{2}m^{2}\cos^{2}[\lambda(m)]}{\chi_{b}(m)}t+
\frac{g_{0}^{2}m^{2}\sin^{2}[\lambda(m)]}{\chi_{c}(m)}t\nonumber\\
&&+\{\eta(m)\text{Im}[\beta_{0}]-\beta(m)\text{Im}[\eta_{0}]\}\sin[\lambda(m)]\nonumber\\
&&+\{\beta(m)\text{Im}[\beta_{0}]+\eta(m)\text{Im}[\eta_{0}]\}\cos[\lambda(m)]\nonumber\\
&&-\beta^{2}(m)\sin[\chi_{b}(m)t]-\eta^{2}(m)\sin[\chi_{c}(m)t]\nonumber\\
&&+\beta(m)\left\{\text{Im}[\eta_{0}e^{-i\chi_{b}(m)t}]\sin[\lambda(m)]\right.\nonumber\\
&&\left. -\text{Im}[\beta_{0}e^{-i\chi_{b}(m) t}]\cos[\lambda(m)]\right\}\nonumber \\
&&-\eta(m)\left\{\text{Im}[\eta_{0}e^{-i\chi_{c}(m)t}]\cos[\lambda(m)]\right.\nonumber\\
&&\left. +\text{Im}[\beta_{0}e^{-i\chi_{c}(m)t}]\sin[\lambda(m)]\right\}.
\end{eqnarray}
%%%%%%%%%%%%%%%%%%%%%%%%%%%%%
\begin{figure}[tbp]
\center\includegraphics[width=0.48\textwidth]{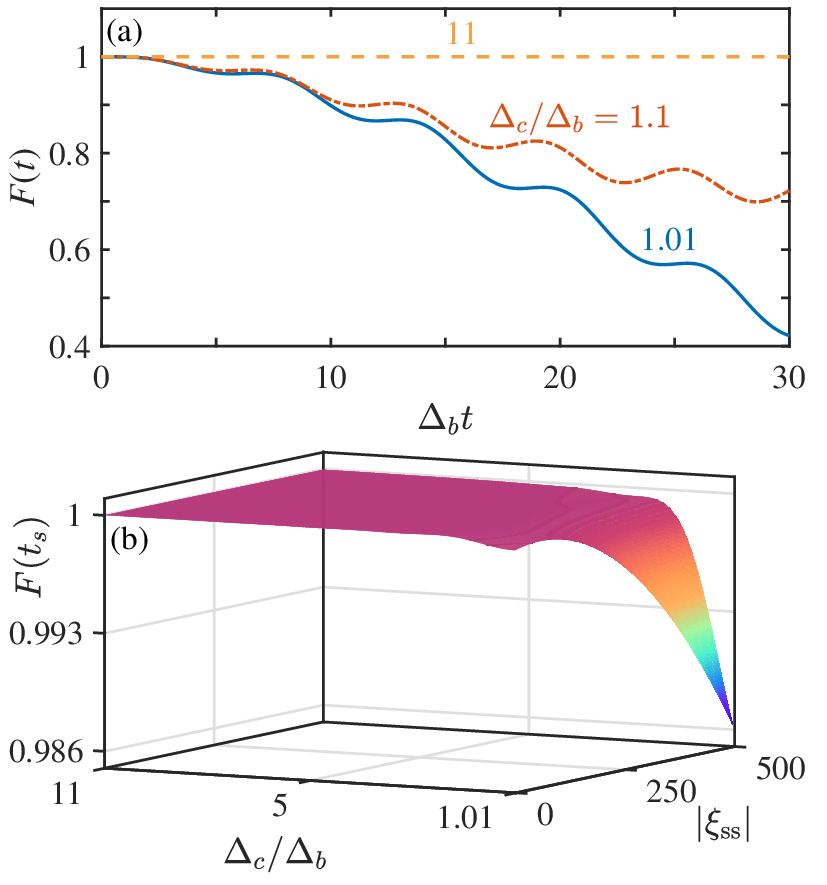}
\caption{(a) Fidelity $F(t)$ defined by Eq.~(\ref{fidclos}) as a function of time $\Delta_{b}t$ when $\Delta_{c}/\Delta_{b}=1.01$, $1.1$, and $11$. (b) Fidelity $F(t_{s})$ as a function of the parameters $\Delta_{c}/\Delta_{b}$ and $\vert\xi_{\text{ss}}\vert$. The other parameters used in both (a) and (b) are $g/\Delta_{b}=0.01$ and $|\xi_{\text{ss}}|=500$, and we choose the initial state $|1\rangle_{a}|\beta_{0}\rangle_{b}|\eta_{0}\rangle_{c}$ of the system with $\beta_{0}=\eta_{0}=0.8$.}
\label{fidvsdeltaandxifig}
\end{figure}
%%%%%%%%%%%%%%%%%%%%%%%%%%%%%

\subsubsection{The fidelity between the approximate state $|\Psi_{\mathrm{app}}(t)\rangle$ and the exact state $|\Psi_{\mathrm{ext}}(t)\rangle$}
Based on the expressions of the exact and approximate states, we can calculate the fidelity between the exact state $|\Psi_{\text{ext}}(t)\rangle$ and the approximate state $|\Psi_{\text{app}}(t)\rangle$ as
\begin{eqnarray}
\label{fidclos}
F(t) &=&|\langle \Psi _{\text{app}}(t)|\Psi _{\text{ext}}(t)\rangle |\nonumber
\\
&=&\left\vert \exp \left\{ -\tfrac{1}{2}\right. [|\beta _{1}(m)|^{2}+|\beta
_{2}(m)|^{2}+|\eta _{1}|^{2}+|\eta _{2}(m)|^{2}]\right.\nonumber \\
&&+\left. \left. \beta _{1}^{\ast }(m)\beta _{2}(m)+\eta _{1}^{\ast }\eta
_{2}(m)\right\} \right\vert .
\end{eqnarray}

Without loss of generality, here we assume the initial state $|\Psi(0)\rangle=|1\rangle_{a}|\beta_{0}\rangle_{b}|\eta_{0}\rangle_{c}$ of the system, with $|\beta_{0}\rangle_{b}$ and $|\eta_{0}\rangle_{c}$ coherent states. Note that the initial time here corresponds to the time when the system reaches its steady state. In addition, we include mode $c$ in $|\psi_{\text{app}}(0)\rangle$ in our simulations. This is because our simulations are performed for the total system, though mode $c$ decouples with modes $a$ and $b$. In particular, we first consider the closed-system case for avoiding the crosstalk from the system dissipation. In Fig.~\ref{fidvsdeltaandxifig}(a) we display the fidelity $F(t)$ given by Eq.~(\ref{fidclos}) as a function of the evolution time $t$ when $\beta_{0}=\eta_{0}=0.8$ and $\Delta_{c}/\Delta_{b}=1.01$, $1.1$, and $11$. In addition, we use the scaled parameters in our numerical simulations for indicating the universality of our scheme. These parameters are within the reach of current experimental conditions.  In Fig.~\ref{fidvsdeltaandxifig}(a), we find that a higher fidelity can be obtained for a larger value of the ratio of $\Delta_{c}/\Delta_{b}$, which coincides with the parameter conditions for the RWA. To show that the fidelity is high in a wide parameter space, we plot the fidelity $F(t_{s})$ at time $t_{s}=\pi/\Delta_{b}$ (the time for generation of the cat state in mode $b$) as a function of the two tunable parameters $\Delta_{c}/\Delta_{b}$ and $\vert\xi_{\text{ss}}\vert$ in Fig.~\ref{fidvsdeltaandxifig}(b). Here the fidelity is high in our selected parameter space and hence the approximate Hamiltonian~(\ref{appHamiltonian}) can properly describe the system. Note that the term $\omega_{a}{a}^{\dagger}{a}$ commutates with other terms in the Hamiltonian, and hence the fidelity is independent of $\omega_{a}$.
%%%%%%%%%%%%%%%%%%%%%%%%%%%%%
\begin{figure*}[tbp]
\center
\includegraphics[ width=0.92 \textwidth]{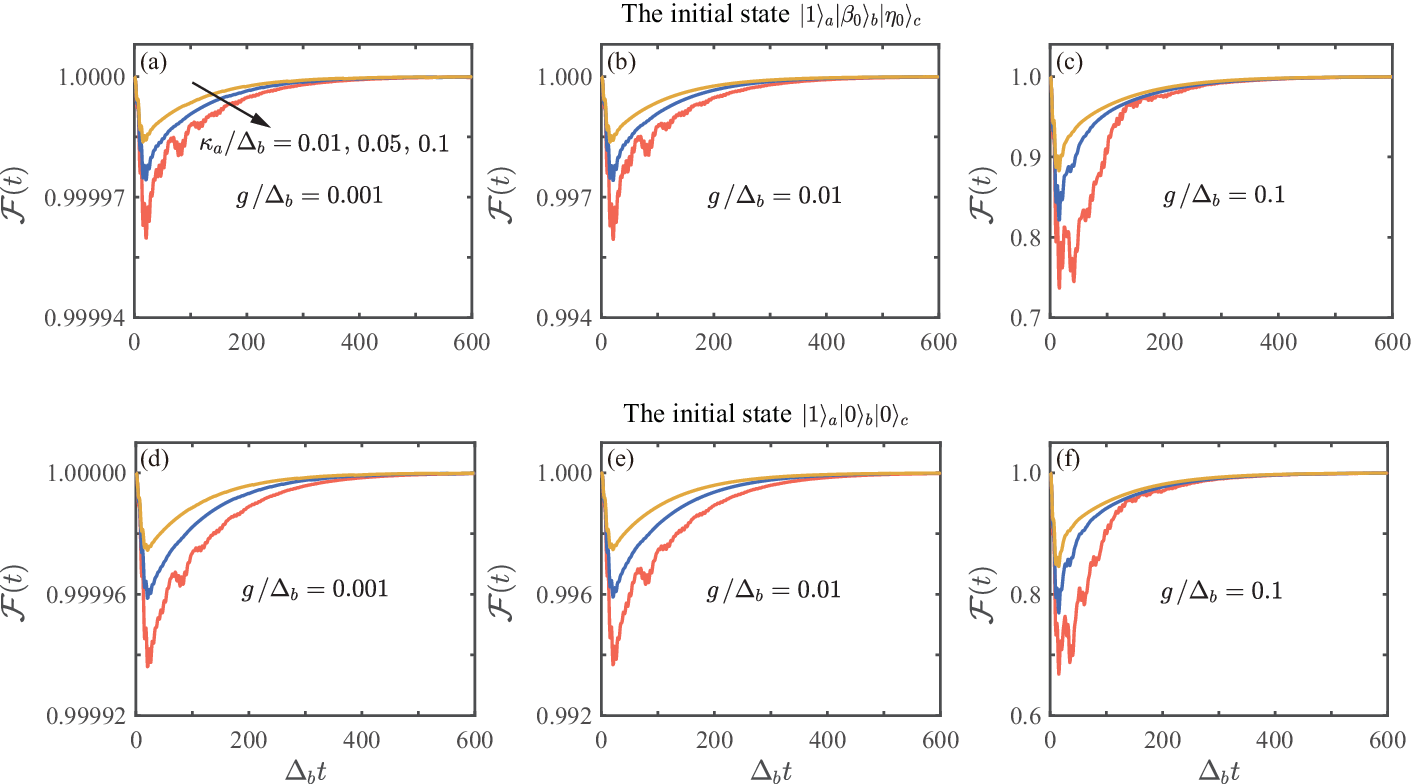}
\caption{Fidelity defined by Eq.~(\ref{fidopen}) versus the evolution time $\Delta_{b}t$, when the parameter $g$ takes various values: (a) and (d) $g/\Delta_{b}=0.001$, (b) and (e) $g/\Delta_{b}=0.01$, and (c) and (f) $g/\Delta_{b}=0.1$. Here we choose a proper parameter $\vert\xi_{\text{ss}}\vert$ such that $g\vert\xi_{\text{ss}}\vert\sim\Delta_{b}$ is satisfied. The initial states of the three modes are chosen as (a)$-$(c) $|1\rangle_{a}|\beta_{0} \rangle_{b}|\eta_{0}\rangle _{c}$ with $\beta_{0}=\eta_{0}=0.2$ and (d)$-$(f) $|1\rangle_{a}|0 \rangle_{b}|0\rangle_{c}$. The other parameters are $\Delta_{c}/\Delta_{b}=1.2$, $\kappa_{b}/\Delta_{b}=\kappa_{c}/\Delta_{b}=0.01$, and $\bar{n}_{o=a,b,c}=0$.}
\label{fidopenfig}
\end{figure*}
%%%%%%%%%%%%%%%%%%%%%%%%%%%%%

We also investigate the influence of the system dissipation on the fidelity. In the open-system case, the evolution of the exact and approximate states are governed by the exact and approximate quantum master equations, respectively. The latter takes the same form as the exact quantum master equation~(\ref{excMasEq}) under the replacement $H_{\text{dis}}\rightarrow H_{\text{app}}$. In addition, for calculational convenience we introduce a parameter $\varepsilon$ into the term $ga^{\dagger}a(b^{\dagger}c+c^{\dagger}b)$ in the displaced Hamiltonian by replacing $g$ $\rightarrow $ $\varepsilon g$; then the exact and approximate equations of motion for the density-matrix elements can be expressed in a unified form. The values of $\varepsilon=1$ and $\varepsilon=0$ correspond to the exact and approximate solution cases, respectively. For convenience, we express the density matrix of the full system in the Fock-state space as
\begin{equation}
\label{MQ of open F}
\rho^{\prime}=\sum_{m,j,s,n,k,r=0}^{\infty}\rho_{m,j,s,n,k,r}^{\prime}|m\rangle_{a}|j\rangle_{b}|s\rangle_{c}\;_{a}\langle n|_{b}\langle
k|_{c}\langle r|,
\end{equation}
with
\begin{equation}
\rho_{m,j,s,n,k,r}^{\prime}=\;_{a}\langle m|_{b}\langle j|_{c}\langle s|\rho^{\prime}|n\rangle _{a}|k\rangle_{b}|r\rangle _{c}.
\end{equation}
In our numerical simulations, the appropriate truncation dimension of the summations in Eq.~(\ref{MQ of open F}) needs to be chosen and Eq.~(\ref{MQ for cat-state}) also follows the same convention. To solve the equations of motion for these density-matrix elements, we assume that the initial state of the system is either $|1\rangle_{a}|\beta_{0} \rangle_{b}|\eta_{0}\rangle _{c}$ with coherent states $|\beta_{0} \rangle_{b}$ and $|\eta_{0}\rangle _{c}$ or $|1\rangle_{a}|0 \rangle_{b}|0\rangle_{c}$; then the corresponding initial conditions are given, respectively by
\begin{subequations}
\label{openinicondition}
\begin{align}
\rho_{m,j,s,n,k,r}^{\prime }(0)&=\delta _{m,1}\delta _{n,1}e^{-\left\vert
\beta \right\vert ^{2}}e^{-\left\vert \gamma \right\vert ^{2}}\frac{\beta
^{j}\beta ^{\ast k}\gamma^{s}\gamma ^{\ast r}}{\sqrt{j!s!k!r!}},\\
\rho _{m,j,s,n,k,r}^{\prime}(0)&=\delta _{m,1}\delta _{j,0}\delta
_{s,0}\delta _{n,1}\delta _{k,0}\delta _{r,0}.
\end{align}
\end{subequations}
With Eq.~(\ref{openinicondition}), the fidelity between the exact and approximate density matrices $\rho_{\text{ext}}$ and $\rho_{\text{app}}$ (corresponding to $\varepsilon=1$ and $\varepsilon=0$, respectively) is given by
\begin{equation}
\mathcal{F}(t)=\text{Tr}\left[\sqrt{\sqrt{\rho_{\text{ext}}}\rho_{\text{app}}\sqrt{\rho_{\text{ext}}}}\right].\label{fidopen}
\end{equation}

In the presence of dissipation, the relaxation time is of the order of $1/\kappa_{o=a,b,c}$, which is much shorter than the time scale $\pi/\Delta_{b}$ due to $1/\kappa_{o=a,b,c}\gg g$. This means that our scheme works in the weak-coupling regime for the initial interaction determined by the term $ga^{\dagger}a(b^{\dagger}c+c^{\dagger}b)$. To show the feasibility of our method, in Fig.~\ref{fidopenfig} we plot the fidelity given by Eq.~(\ref{fidopen}) as a function of the evolution time $\Delta_{b}t$ in the open-system case. In principle, the initial state of the system can be chosen arbitrarily, whereas to save computational resources, we select the initial state of the system as either $|1\rangle_{a}|\beta_{0}\rangle_{b}|\eta_{0}\rangle_{c}$ or $|1\rangle_{a}|0 \rangle_{b}|0\rangle_{c}$. Meanwhile, for realization of the ultrastrong-coupling regime, a proper displacement amplitude $\vert\xi_{\text{ss}}\vert$ is considered so that the relation $g\vert\xi_{\text{ss}}\vert\sim\Delta_{b}$ can be satisfied. Here we find that a smaller value of the ratio $g/\Delta_{b}$ corresponds to a higher fidelity, which confirms our analysis for the parameter conditions of the RWA. In addition, the fidelity exhibits some oscillations and then reaches gradually a stationary value due to the system dissipation. For a given $g$, corresponding to the higher decay rate of the system, the faster the fidelity decays to a stationary value. However, from Fig.~\ref{fidopenfig} we see that the values of the fidelities increase gradually over time until they are equal to 1. The physical origin can be seen from the approximate Hamiltonian $H_{\text{app}}$ and the exact Hamiltonian $H_{\text{dis}}$, in which mode $a$ is not driven but couples to a zero-temperature reservoir. Therefore, the average photon number in mode $a$ becomes zero in the long-time limit, namely, obtaining the steady state $|0\rangle_{a}$ of mode $a$. This vacuum state of mode $a$ can lead to the disappearance of the terms $ga^{\dagger}a(b^{\dagger}c+c^{\dagger}b)$ and $ga^{\dagger}a(\xi b^{\dagger}+\xi b)$; then both modes $b$ and $c$ are reduced to free-cavity modes connected with individual zero-temperature  reservoirs. Hence the corresponding steady states of modes $b$ and $c$ are $|0\rangle_{b}$ and $|0\rangle_{c}$, respectively. As a result, the fidelity between exact and approximate states is one in the long-time limit.
%%%%%%%%%%%%%%%%%%%%%%%%%%%%%
\begin{figure}[tbp]
\center\includegraphics[width=0.48\textwidth]{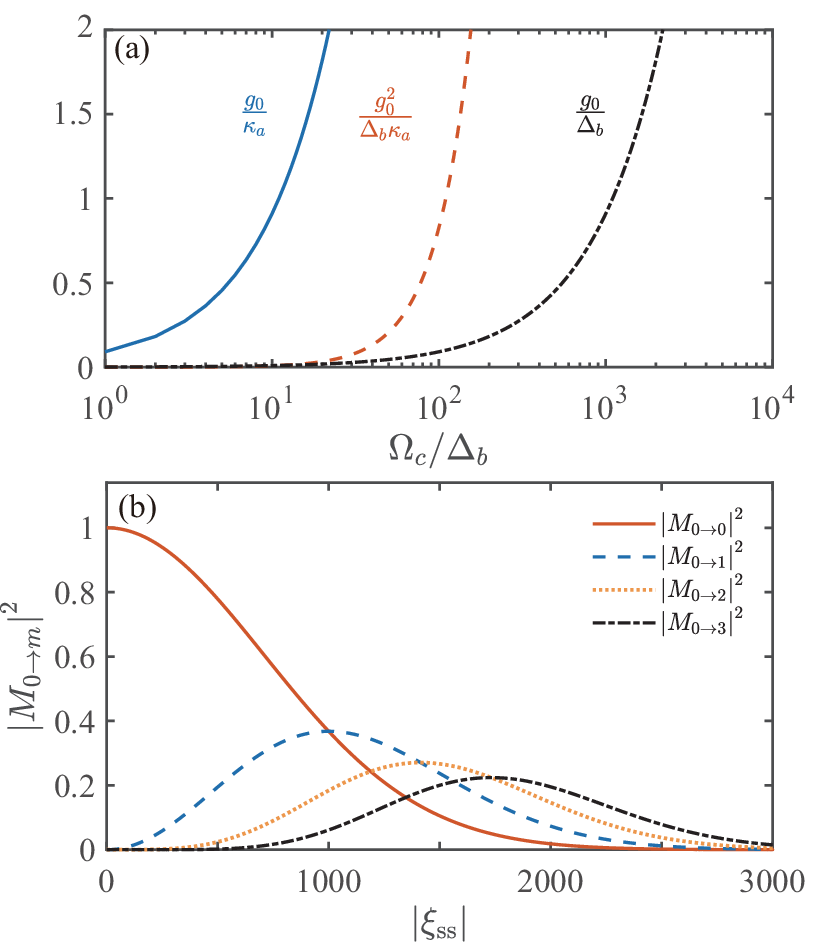}
\caption{(a) Three important ratios $g_{0}/\kappa_{a}$, $g_{0}^{2}/\Delta_{b}\kappa_{a}$, and $g_{0}/\Delta_{b}$ versus the driving amplitude $\Omega_{c}/\Delta_{b}$. The other parameters are $g/\Delta_{b}=0.001$, $\Delta_{c}/\Delta_{b}=1.1$, $\kappa_{a}/\Delta_{b}=0.01$, and $\kappa_{c}/\Delta_{b}=0.001$. (b) The FC factors for transitions $|0\rangle_{a}|0\rangle_{b}\leftrightarrow |1\rangle _{a}|\tilde{m}(1)\rangle_{b}$ versus the amplification factor $\left\vert\xi_{\text{ss}}\right\vert$ with $m=0$, 1, 2, and 3. The other parameters are $g/\Delta_{b}=0.001$ and $\Delta_{c}/\Delta_{b}=20$.}
\label{parameteranalysisfig}
\end{figure}
%%%%%%%%%%%%%%%%%%%%%%%%%%%%%

\subsubsection{Parameter space analysis}
To clearly see the coupling enhancement in this system, we present an analysis of the parameter space of the enhanced optomechanical interaction. Concretely, we analyze three ratios, $g_{0}/\kappa_{a}$, $g_{0}^{2}/\Delta_{b}\kappa_{a}$, and $g_{0}/\Delta_{b}$, which have important physical meaning in optomechanical systems~\cite{Aspelmeyer2014}. In optomechanics, $g_{0}/\kappa_{a}>1$ corresponds to the single-photon strong-coupling regime, in which the mechanical quantum fluctuation induced by a single photon can be resolved from the zero-point fluctuation of the mechanical mode~\cite{Ludwig2008}. The relation $g_{0}^{2}/\Delta_{b}\kappa_{a}>1$ represents the strong-dispersive-coupling condition which ensures that the energy nonharmonicity induced by the Kerr nonlinearity can be resolved from the cavity emission spectrum of mode $a$ in the case of $g_{0}\ll \Delta_{b}$~\cite{Rabl2011}. In addition, the relation $g_{0}/\Delta_{b}>1$ denotes the deep-strong-coupling condition which means that the displacement of mode $b$ forced by a single photon can be distinguished from the vacuum state of mode $b$~\cite{Marshall2003}.
In Fig.~\ref{parameteranalysisfig}(a) we show the ratios $g_{0}/\kappa_{a}$, $g_{0}^{2}/\Delta_{b}\kappa_{a}$, and $g_{0}/\Delta_{b}$ as functions of the driving amplitude $\Omega_{c}/\Delta_{b}$. Here we can see that for a large driving amplitude, these three ratios can be larger than 1, which means that single-photon optomechanical effects can be observed in this system.

The enhanced optomechanical interaction can also be witnessed by analyzing the transition suppression effect. We consider the transitions between the states $|0\rangle_{a}|0\rangle_{b}$ and $|1\rangle _{a}|\tilde{m}(1)\rangle_{b}$, where $|\tilde{m}(1)\rangle_{b}=e^{\beta(1)(b^{\dagger}-b)}|m\rangle_{b}$ defines the single-photon displaced number state with $\beta(1)$ the single-photon displacement depending on $\vert\xi_{\text{ss}}\vert$ [see Eq.~(\ref{beta1}) for the expression of $\beta(1)$]. The transition probability, called the Franck-Condon (FC) factor~\cite{Franck1925,Condon1926,Leturcq2009}, is proportional to the square of the overlap between the ground state $|0\rangle_{b}$ and the displaced number states $|\tilde{m}(1)\rangle_{b}$, and it takes the form
\begin{equation}
|M_{0\rightarrow m}|^{2}=\left\vert e^{-|\beta (1)|^{2}/2}\frac{[-\beta(1)]^{m}}{\sqrt{m!}}\right\vert ^{2}.
\end{equation}
Figure~\ref{parameteranalysisfig}(b) shows the FC factors as a function of the enhanced factor $\left\vert\xi_{\text{ss}}\right\vert$, from which we see that the ground-state-to-ground-state transition, determined by $\left\vert M_{0\rightarrow 0}\right\vert^{2}$, is exponentially suppressed for a large $\left\vert\xi_{\text{ss}}\right\vert$. However, the peak values of other FC factors $|M_{0\rightarrow m}|^{2}$ are located at a larger $|\xi_{\text{ss}}|$ for a larger $m$. Furthermore, the peak values of these FC factors decrease gradually as the excited number $m$ increases.

\section{Macroscopic quantum superposition}\label{Cat-state generation}
In this section we show how to generate the Schr\"{o}dinger cat states for the mechanical-like mode $b$~\cite{Marshall2003} in terms of the approximate Hamiltonian $H_{\text{app}}$ and the exact Hamiltonian $H_{\text{dis}}$. We also study the influence of the system dissipation on the cat-state generation.

\subsection{Cat-state generation based on the approximate Hamiltonian $H_{\mathrm{app}}$}
In a rotating frame with respect to $H_{0}=\omega_{a}a^{\dagger}a+\Delta_{b}b^{\dagger}b+\Delta_{c}c^{\dagger}c$, the approximate Hamiltonian $H_{\text{app}}$ becomes
\begin{equation}
H_{\text{app}}^{(I)}(t)=-g_{0}a^{\dagger}a(b^{\dagger}e^{i\theta_{c}}e^{i\Delta_{b}t}+be^{-i\theta_{c}}e^{-i\Delta_{b}t}).
\end{equation}
The unitary evolution operator associated with $H_{\text{app}}$ is given by
\begin{equation}
\label{appunitaryop}
U_{\text{app}}(t)=e^{-iH_{0}t}U^{(I)}_{\text{app}}(t),
\end{equation}
where $U^{(I)}_{\text{app}}(t)$ is the unitary evolution operator relevant to $H_{\text{app}}^{(I)}(t)$ and it is governed by the equation of motion $i\partial U_{\text{app}}^{(I)}(t)/\partial t=H_{\text{app}}^{(I)}(t)U_{\text{app}}^{(I)}(t)$ with initial condition given by $U^{(I)}_{\text{app}}(0)=I$. Its formal solution can be written as $U^{(I)}_{\text{app}}(t)=\mathcal{T}\exp{[-i\int_{0}^{t}H_{\text{app}}^{(I)}(t^{\prime})dt^{\prime}]}$, where $\mathcal{T}$ denotes the time-ordering operator. According to the Magnus proposal, $U^{(I)}_{\text{app}}(t)$ can be obtained as
\begin{eqnarray}
\label{the propagator}
U_{\text{app}}^{(I)}(t)&=&\exp\left\{i\frac{g_{0}^{2}}{\Delta_{b}^{2}}[\Delta_{b}t\!-\!\sin(\Delta_{b}t)]a^{\dagger }aa^{\dagger}a\right\}\nonumber\\
&&\times\exp\left\{\frac{g_{0}}{\Delta_{b}}a^{\dagger}a[b^{\dagger}(e^{i\Delta_{b}t}\!-\!1)e^{i\theta _{c}}\!-\!b(e^{-i\Delta _{b}t}\!-\!1)e^{-i\theta_{c}}]\right\}.\nonumber\\
&&
\end{eqnarray}

\begin{figure}[tbp]
\center
\includegraphics[ width=0.48 \textwidth]{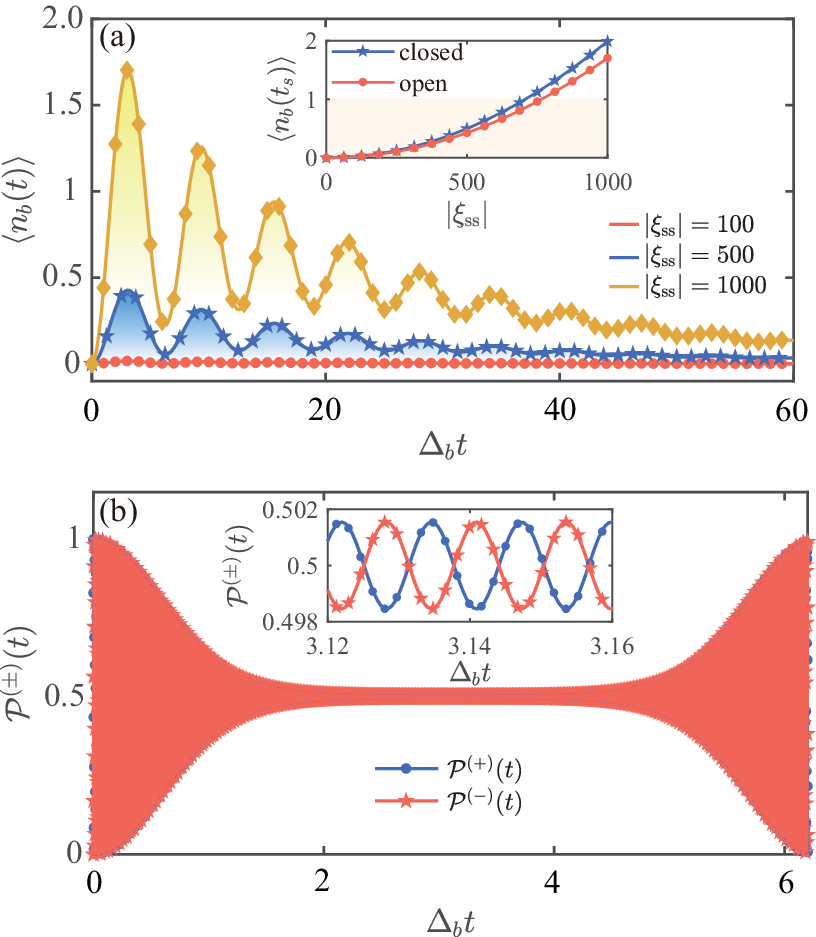}
\caption{(a) Dynamics of the average excitation number $\langle n_{b}(t)\rangle$ of mode $b$ at different values of the enhanced factor: $\vert\xi_{\text{ss}}\vert=100$, $500$, and $1000$. The inset shows the average excitation number $\langle n_{b}(t_s)\rangle$ in the closed- and open-system cases at time $t_s=\pi/|\Delta_{b}|$ as a function of $\vert\xi_{\text{ss}}\vert$. The other parameters are $g/\Delta_{b}=0.001$, $\Delta_{c}/\Delta_{b}=20$, $\kappa_{o=a,b,c}/\Delta_{b}=0.05$, and $\bar{n}_{o=a,b,c}=0$. (b) Time dependence of the detection probabilities $\mathcal{P}^{(\pm)}(t)$. The inset in (b) is a close-up of $\mathcal{P}^{(\pm)}(t)$ in the middle duration of one period. The other parameters are $\omega_{a}/\Delta_{b}=500$, $\Delta_{c}/\Delta_{b}=20$, $g/\Delta_{b}=0.001$, and $\vert\xi_{\text{ss}}\vert=1700$. For all plots in both (a) and (b), the markers represent the results obtained based on the approximate Hamiltonian $H_{\text{app}}$ and the solid curves correspond to the results associated with the exact Hamiltonian $H_{\text{dis}}$.}
\label{averexcitnumberfig}
\end{figure}

To generate the cat states, we choose the initial state of the system as
\begin{equation}
|\psi_{\text{app}}(0)\rangle=\frac{1}{\sqrt{2}}(|0\rangle_{a}+|1\rangle_{a})|0\rangle_{b}|0\rangle_{c},
\end{equation}
where we select the initial state of mode $c$ as $|0\rangle_{c}$ to save computational resources. It should be pointed out that the initial state of mode $c$ in the original representation is a coherent state with the coherent amplitude given by $\xi_{\text{ss}}=\Omega_{c}/(\Delta_{c}-i\kappa_{c}/2)$. Therefore, according to the unitary evolution operator~(\ref{appunitaryop}), the state of the system at time $t$ can be obtained as
\begin{equation}
\label{ana state at t}
|\psi_{\text{app}}(t)\rangle=\frac{1}{\sqrt{2}}[|0\rangle_{a}|0\rangle_{b}+e^{i\varphi(t)}
|1\rangle_{a}|\beta(t)\rangle_{b}]|0\rangle_{c},
\end{equation}
where we introduce the phase
\begin{equation}
\label{thetaandeta}
\varphi(t)=\frac{g_{0}^{2}}{\Delta_{b}^{2}}[\Delta_{b}t-\sin(\Delta_{b}t)]-\omega_{a}t
\end{equation}
and the displacement amplitude
\begin{equation}
\label{thetaandeta}
\beta(t)=\frac{g_{0}}{\Delta_{b}}(1-e^{-i\Delta_{b}t})e^{i\theta_{c}}.
\end{equation}
From Eq.~(\ref{thetaandeta}) we see that the maximal displacement $\vert\beta\vert_{\max }=2g_{0}/|\Delta_{b}|$ is obtained at time $t=(2n+1)\pi/|\Delta_{b}|$ for natural numbers $n$. To see the displacement effect in mode $b$ induced by a single photon, we calculate the average excitation $\langle n_{b}(t)\rangle=\langle b^{\dagger}b\rangle=|\beta(t)|^{2}/2$ in mode $b$ based on the approximate Hamiltonian $H_{\text{app}}$. As shown by the markers in Fig.~\ref{averexcitnumberfig}(a), a larger $\vert\xi_{\text{ss}}\vert$ will lead to a larger maximal displacement, and the dissipation will decrease the maximal value of the displacement. In particular, we plot the average excitation number $\langle n_{b}(t_s)\rangle$ at the cat-state generation time $t_{s}=\pi/|\Delta_{b}|$ as a function of $\vert\xi_{\text{ss}}\vert$ [see the markers in the inset of Fig.~\ref{averexcitnumberfig}(a)]. Here we can see that the peak value of the displacement could be larger than the zero-point fluctuations of mode $b$ in the presence of dissipation.
%%%%%%%%%%%%%%%%%%%%%%%%%%%%%
%%%%%%%%%%%%%%%%%%%%%%%%%%%%%

%%%%%%%%%%%%%%%%%%%%%%%%%%%%%
\begin{figure*}[tbp]
\center
\includegraphics[width=0.86 \textwidth]{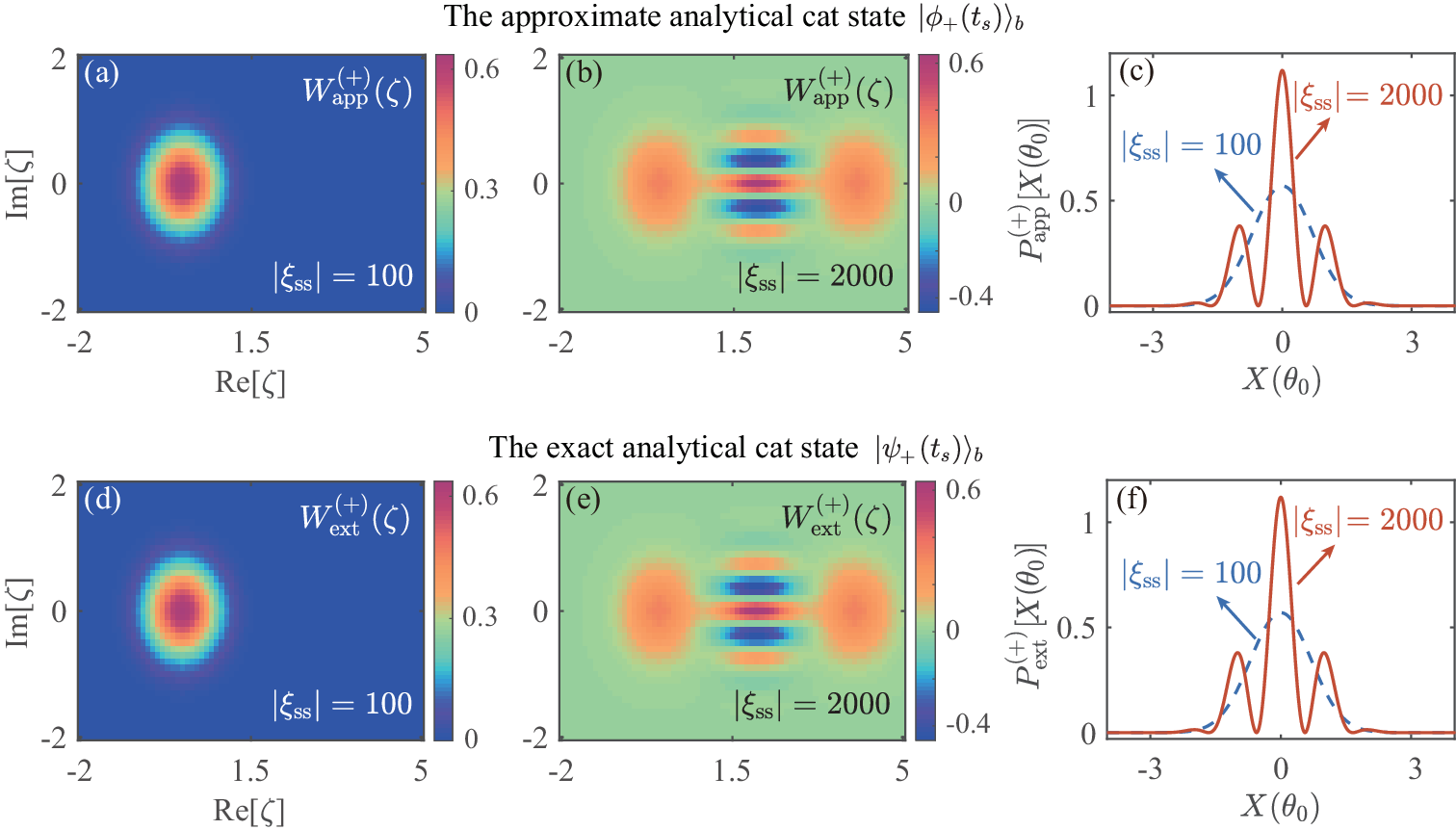}
\caption{Wigner functions (a) and (b) $W_{\text{app}}^{(+)}(\zeta)$ of the approximate analytical cat state $|\phi_{+}(t_{s})\rangle_{b}$ and (d) and (e) $W_{\text{ext}}^{(+)}(\zeta)$ of the exact analytical cat state $|\psi_{+}(t_{s})\rangle_{b}$. Here the amplification factor $|\xi_{\text{ss}}|$ takes different values: (a) and (d) $|\xi_{\text{ss}}|=100$ and (b) and (e) $|\xi_{\text{ss}}|=2000$. (c) and (f) Probability distributions $P_{\text{app}}^{(+)}[X(\theta_{0})]$ for the states
$|\phi_{+}(t_{s})\rangle_{b}$ and $P_{\text{ext}}^{(+)}[X(\theta_{0})]$ for $|\psi_{+}(t_{s})\rangle_{b}$ versus $X(\theta_{0})$ at different values of $|\xi_{\text{ss}}|$. The other parameters are $g/\Delta_{b}=0.001$, $\Delta_{c}/\Delta_{b}=20$, $t_{s}=\pi/|\Delta_{b}|$, and $\theta_{0}=\arg[\beta(t_{s})]-\pi/2$.}
\label{Wignerclosfig}
\end{figure*}
%%%%%%%%%%%%%%%%%%%%%%%%%%%%%

To generate the macroscopic quantum superposed states of mode $b$, we measure the state of modes $a$ and $c$ with the basis states $|\pm\rangle_{a}|0\rangle_{c}=(|0\rangle_{a}\pm|1\rangle_{a})|0\rangle_{c}/\sqrt{2}$. In terms of the basis states $|\pm\rangle_{a}|0\rangle_{c}$, the state in Eq.~(\ref{ana state at t}) can be reexpressed as
\begin{eqnarray}
|\psi _{\text{app}}(t)\rangle  &=&\frac{1}{2}\left\{ |+\rangle _{a}|0\rangle
_{c}[|0\rangle _{b}+e^{i\varphi (t)}|\beta (t)\rangle _{b}]\right.   \nonumber
\\
&&+\left. |-\rangle _{a}|0\rangle _{c}[|0\rangle _{b}-e^{i\varphi (t)}|\beta
(t)\rangle _{b}]\right\}.
\end{eqnarray}
If we perform a measurement on the system with the basis states $|\pm\rangle_{a}|0\rangle_{c}$, mode $b$ will collapse into two analytical cat states
\begin{equation}
\label{analytical cat states}
|\phi_{\pm}(t)\rangle_{b}=\mathcal{N}_{\pm}[|0\rangle_{b}\pm e^{i\varphi(t)}|\beta(t)\rangle_{b}],
\end{equation}
where the normalization constants are defined by
\begin{equation}
\label{normalization constant}
\mathcal{N}_{\pm }=\{2[1\pm e^{-|\beta (t)|^{2}/2}\cos \varphi (t)]\}^{-1/2}.
\end{equation}
The measuring probabilities corresponding to states $|\pm\rangle_{a}|0\rangle_{c}$ are given by
\begin{equation}
\label{analytical probabilities}
\mathcal{P}_{\text{app}}^{(\pm )}(t)=\frac{1}{2}[1\pm e^{-|\beta(t)|^{2}/2}\cos \varphi (t)].
\end{equation}
Equation~(\ref{analytical probabilities}) indicates that, for a sufficiently large displacement $\vert\beta(t)\vert$, the measurement probabilities $\mathcal{P}^{(\pm)}_{\text{app}}(t)$ will approach $\frac{1}{2}$ due to $\exp[-\left\vert\beta(t)\right\vert^{2}/2]\approx 0$. In Fig.~\ref{averexcitnumberfig}(b) we use the markers to denote the probabilities $\mathcal{P}^{(\pm)}_{\text{app}}(t)$ as functions of the evolution time $\Delta_{b}t$, which shows that $\mathcal{P}^{(+)}_{\text{app}}(t)$ and $\mathcal{P}^{(-)}_{\text{app}}(t)$ have similar oscillation envelops. In addition, the inset in Fig.~\ref{averexcitnumberfig}(b) indicates that the oscillation amplitude is almost negligible and the probabilities $\mathcal{P}^{(+)}_{\text{app}}(t)\approx\mathcal{P}^{(-)}_{\text{app}}(t)\approx\frac{1}{2}$ in the intermediate duration around the detection time $t_{s}=\pi/|\Delta_{b}|\approx3.14$, which confirms our analysis of the measurement probabilities $\mathcal{P}^{(\pm)}_{\text{app}}(t)$ [denoted by the markers in Fig.~\ref{averexcitnumberfig}(b)].

To see the quantum interference and coherence effects in the generated Schr\"{o}dinger cat states, we now calculate the Wigner function. For a single-mode system described by the density matrix $\rho$, the Wigner function is defined by~\cite{Barnettbook}
\begin{equation}
\label{the def of W finction}
W(\zeta)=\frac{2}{\pi}\mathrm{Tr}\left[D^{\dagger}(\zeta)\rho D(\zeta)(-1)^{b^{\dagger}b}\right],
\end{equation}
where $D(\zeta)=\exp(\zeta b^{\dagger}-\zeta^{\ast}b)$ is the displacement operator. Corresponding to the approximate analytical cat states $|\phi_{\pm}(t)\rangle_{b}$ in Eq.~(\ref{analytical cat states}), the Wigner functions can be obtained by substituting the density matrices $\rho_{\pm}=|\phi_{\pm}\left(t\right)\rangle_{bb}\langle\phi_{\pm}\left(t\right)|$ into Eq.~(\ref{the def of W finction}) as
\begin{eqnarray}
W_{\text{app}}^{(\pm )}(\zeta ) &=&\frac{2|\mathcal{N}_{\pm }|^{2}}{\pi }
\left( e^{-2\left\vert \zeta \right\vert ^{2}}+e^{-2\left\vert \zeta -\beta
\left( t\right) \right\vert ^{2}}\right.   \nonumber \\
&&\pm 2\text{Re}\left[ e^{-i\varphi \left( t\right) }e^{i\text{Im}[\zeta
\beta ^{\ast }\left( t\right) ]}\right.   \nonumber \\
&&\left. \left. \times e^{-2\left\vert \zeta \right\vert ^{2}+(1/2)\zeta
^{\ast }\beta \left( t\right) +(3/2)\zeta \beta ^{\ast }\left( t\right)
-(1/2)\left\vert \beta \left( t\right) \right\vert ^{2}}\right] \right) ,
\end{eqnarray}
with the normalization constants $\mathcal{N}_{\pm}$ given by Eq.~(\ref{normalization constant}). In Figs.~\ref{Wignerclosfig}(a) and~\ref{Wignerclosfig}(b) we plot the Wigner functions $W_{\text{app}}^{(+)}(\zeta)$ for the approximate analytical cat states $|\phi_{+}(t)\rangle_{b}$ at the detection time $t_{s}=\pi/|\Delta_{b}|$ when the dimensionless displacement amplitude $|\xi_{\text{ss}}|=100$ and $|\xi_{\text{ss}}|=2000$, respectively. Here we only plot the Wigner function $W_{\text{app}}^{(+)}(\zeta)$ for concision. By comparison, we see that the magnitude of $|\xi_{\text{ss}}|$ can enhance the distinguishability between the two coherent states $|0\rangle_{b}$ and $|\beta(t)\rangle_{b}$ and the visibility of the interference fringes (in the region between the two peaks). Therefore the enhanced optomechanical coupling realized by our scheme is useful to create macroscopically distinct superposition states in mode $b$.
%%%%%%%%%%%%%%%%%%%%%%%%%%%%%
\begin{figure}[tbp]
\center
\includegraphics[width=0.48\textwidth]{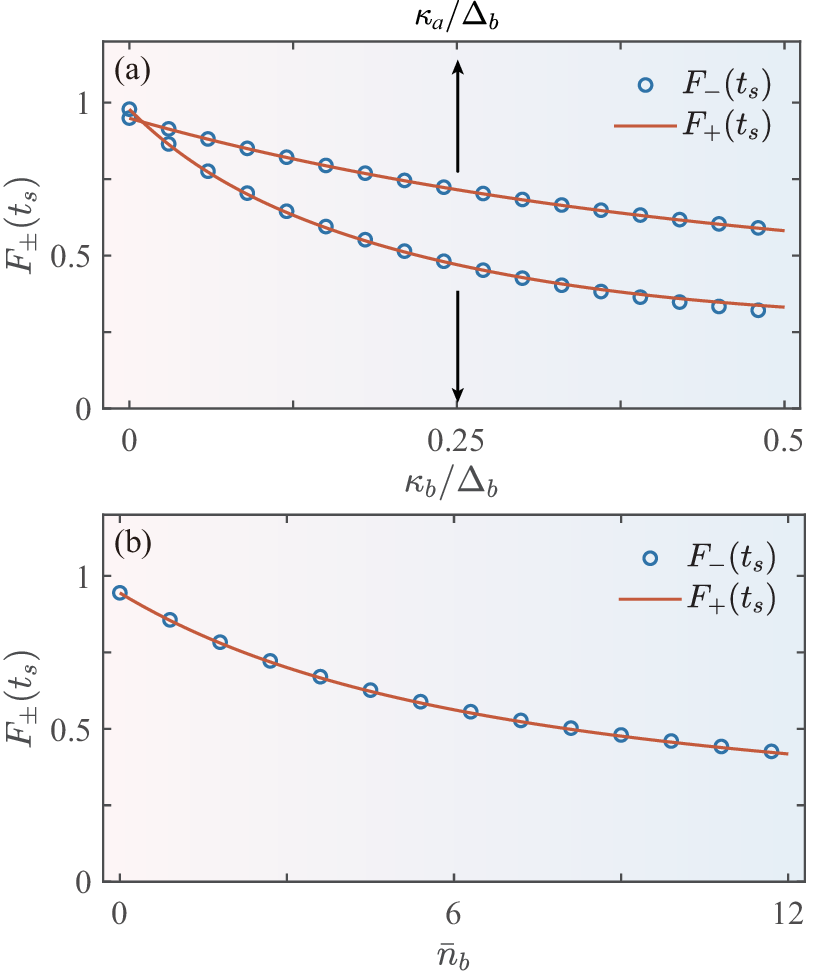}
\caption{Fidelities $F_{\pm}(t_{s})$ versus either $\kappa_{a}/\Delta_{b}$ (the top horizontal axis) or $\kappa_{b}/\Delta_{b}$ (the bottom horizontal axis) at $\kappa_{b}/\Delta_{b}=0.01$ and $\kappa_{a}/\Delta_{b}=0.01$. The other parameters are $\kappa_{c}/\Delta_{b}=0.01$ and $\bar{n}_{o=a,b,c}=0$. (b) Fidelities $F_{\pm}(t_{s})$ versus $\bar{n}_{b}$ at $\kappa_{o=a,b,c}/\Delta_{b}=0.01$ and $\bar{n}_{a}=\bar{n}_{c}=0$.}
\label{fidpmopenfig}
\end{figure}
%%%%%%%%%%%%%%%%%%%%%%%%%%%%%

An alternative approach to observe the quantum coherence and inference effects in the generated cat states is to investigate the probability distribution of the rotated quadrature operator. For the rotated quadrature operator
\begin{equation}
\label{the def of rotqudr operator}
\hat{X}(\theta)=\frac{1}{\sqrt{2}}(be^{-i\theta }+b^{\dagger }e^{i\theta }),
\end{equation}
we can denote its eigenstate by $|X(\theta)\rangle_{b}$: $\hat{X}(\theta)|X(\theta)\rangle_{b}=X(\theta)|X(\theta)\rangle_{b}$~\cite{Milburnbook}. Considering the states $|\phi_{\pm}(t)\rangle_{b}$, we obtain the probability distribution of the rotated quadrature operator $|X(\theta)\rangle_{b}$ as
\begin{eqnarray}
P_{\text{app}}^{(\pm )}[X(\theta )] &=&|\,_{b}\langle X(\theta )|\phi _{\pm
}(t)\rangle _{b}|^{2} \nonumber\\
&=&\mathcal{N}_{\pm }^{2}|\,_{b}\langle X(\theta )|0\rangle _{b}\pm
e^{i\varphi \left( t\right) }\,_{b}\langle X(\theta )|\beta \left( t\right)
\rangle _{b}|^{2},
\end{eqnarray}
with the inner product $_{b}\langle X(\theta)|0\rangle_{b}$ and $_{b}\langle X(\theta )|\beta \left( t\right) \rangle _{b}$ given by
\begin{eqnarray}
\label{relation for rorqudr}
_{b}\langle X(\theta )|0\rangle _{b}\! &=&\!\frac{H_{0}[X(\theta )]}{\sqrt{%
\pi ^{1/2}}}e^{-X^{2}\left( \theta \right) /2},\nonumber\\
_{b}\langle X(\theta )|\beta \left( t\right) \rangle _{b}\!
&=&\!e^{-\left\vert \beta \left( t\right) \right\vert
^{2}/2}\!\sum_{n=0}^{\infty }\frac{[\beta \left( t\right)
]^{n}H_{n}[X(\theta )]}{n!\sqrt{\pi ^{1/2}2^{n}}}e^{-X^{2}\left( \theta
\right) /2}e^{-i\theta n},\nonumber\\
&&
\end{eqnarray}
where $H_{n}[z]$ are the Hermite polynomials.
%%%%%%%%%%%%%%%%%%%%%%%%%%%%%
\begin{figure*}[tbp]
\center
\includegraphics[width=0.98 \textwidth]{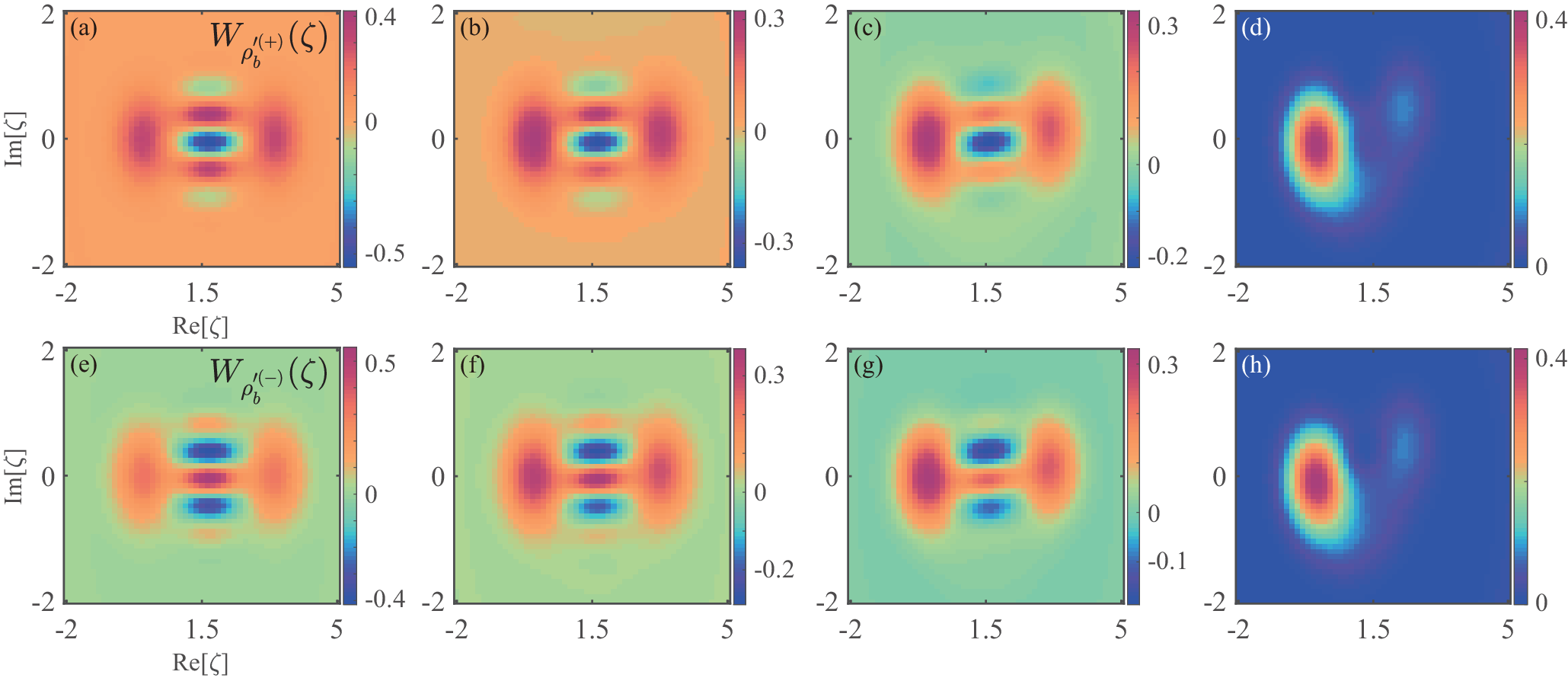}
\caption{Plots of the Wigner functions $W_{\rho_{b}^{\prime(\pm)}}(\zeta)$ of the generated states $\rho_{b}^{\prime(\pm)}(t_{s})$ at selected decay rates: (a) and (e) $\kappa_{o=a,b,c}/\Delta _{b}=0.01$, (b) and (f) $\kappa_{o=a,b,c}/\Delta _{b}=0.05$, (c) and (g) $\kappa_{o=a,b,c}/\Delta _{b}=0.1$, and (d) and (h) $\kappa_{o=a,b,c}/\Delta _{b}=0.5$. The other parameters are $g/\Delta_{b}=0.001$, $\Delta_{c}/\Delta_{b}=20$, $\vert\xi_{\text{ss}}\vert=1700$, $\bar{n}_{o=a,b,c}=0$, and $t_{s}=\pi/|\Delta_{b}|$.}
\label{Wigneropenvsdecayfig}
\end{figure*}
%%%%%%%%%%%%%%%%%%%%%%%%%%%%%

In Fig.~\ref{Wignerclosfig}(c) we plot the probability distributions $P_{\text{app}}^{(+)}[X(\theta_{0})]$ for the state $|\phi_{+}(t)\rangle_{b}$. Here we take the rotated angle $\theta_{0}=\arg[\beta(t_{s})]-\pi/2$; the quadrature direction is, in this case, perpendicular to the link line between the two main peaks. Whenever the two coherent states are projected onto this quadrature, the probability distributions will overlap exactly, which will make the interference maximum. As shown in Fig.~\ref{Wignerclosfig}(c), the larger $\vert\xi_{\text{ss}}\vert$ can cause a stronger oscillation in the probability distributions corresponding to the generated cat states. Note that the probability distributions are much easier to detect than the Wigner functions because the quadrature operators only need to be measured at a given rotating angle.

\subsection{Cat-state generation based on the exact Hamiltonian $H_{\mathrm{ext}}$}
It should be pointed out that the generated cat states can also be calculated based on the exact Hamiltonian $H_{\text{dis}}$. For the given initial state $|\psi_{\text{ext}}(0)\rangle=|\psi_{\text{app}}(0)\rangle$, the exact state at time $t$ can be obtained as
\begin{equation}
\label{anaextstate}
|\psi _{\text{ext}}\! \left( t\right) \rangle \!=\! \frac{1}{\sqrt{2}}
\{ \left \vert 0\right \rangle _{a}\left \vert 0\right \rangle _{b}\left \vert
0\right \rangle _{c}+e^{i\Theta _{\text{ext}}^{(1)}\left( t\right)
}\left \vert 1\right \rangle _{a}\left \vert \beta _{2}(1)\right \rangle
_{b}\left \vert \eta _{2}(1)\right \rangle _{c}\}
\end{equation}
by using the result given in Eq.~(\ref{exact state at t}), where $\beta_{2}(1)$, $\eta_{2}(1)$, and $\Theta_{\text{ext}}^{(1)}(t)$ are given by Eqs.~(\ref{prime beta and prime eta}) and~(\ref{ext phase angle}) at $m=1$, respectively. In terms of the basis states $|\pm\rangle_{a}|0\rangle_{c}$ and Eq.~(\ref{anaextstate}), the exact analytical cat states for mode $b$ after the measurement of modes $a$ and $c$ can be obtained as
\begin{equation}
|\psi _{\pm }(t)\rangle _{b}=\mathcal{K}_{\pm }\{ \left \vert 0\right \rangle
_{b}\pm e^{i\Theta _{\text{ext}}^{(1)}\left( t\right) }e^{-|\eta
_{2}(1)|^{2}/2}\left \vert \beta _{2}(1)\right \rangle _{b}\},
\end{equation}
where the normalization constants are defined by
\begin{equation}
\mathcal{K}_{\pm }=\{1+e^{-|\eta_{2}(1)|^{2}}\pm2 e^{-[|\beta_{2}(1)|^{2}+|\eta_{2}(1)|^{2}]/2}\cos[\Theta_{\text{ext}}^{(1)}(t)]\}^{-1/2}.
\end{equation}
The corresponding probabilities for the measured states $|\pm\rangle_{a}|0\rangle_{c}$ are given by
\begin{equation}
\mathcal{P}_{\text{ext}}^{(\pm )}(t)=\frac{1}{2}\{ 1+e^{-|\eta_{2}(1)|^{2}}\pm 2e^{-[|\beta _{2}(1)|^{2}+|\eta _{2}(1)|^{2}]/2}\cos
[\Theta _{\text{ext}}^{(1)}(t)]\}.
\end{equation}
Based on the above discussion, in Fig.~\ref{averexcitnumberfig}(a) we also plot the average excitation number $\langle n_{b}(t)\rangle$ of mode $b$ as a function of the evolution time $\Delta_{b}t$ when the amplification factor $|\xi_{\text{ss}}|$ takes different values: $|\xi_{\text{ss}}|=100$, $500$, and $1000$ [see the solid curves in Fig.~\ref{averexcitnumberfig}(a)]. The solid curves in the inset of Fig.~\ref{averexcitnumberfig}(a) show the average excitation $\langle n_{b}(t_s)\rangle$ at time $t_{s}=\pi/|\Delta_{b}|$ as a function of $|\xi_{\text{ss}}|$ in both closed- and open-system cases. In addition, the time dependence of the probabilities $\mathcal{P}_{\text{ext}}^{(\pm)}(t)$ is plotted by the solid curves in Fig.~\ref{averexcitnumberfig}(b). We see from Fig.~\ref{averexcitnumberfig} that the results based on the approximate Hamiltonian and the exact Hamiltonian match well, which further confirms the validity of the approximate Hamiltonian $H_{\text{app}}$.
%%%%%%%%%%%%%%%%%%%%%%%%%%%%%
\begin{figure*}[tbp]
\center
\includegraphics[width=0.98 \textwidth]{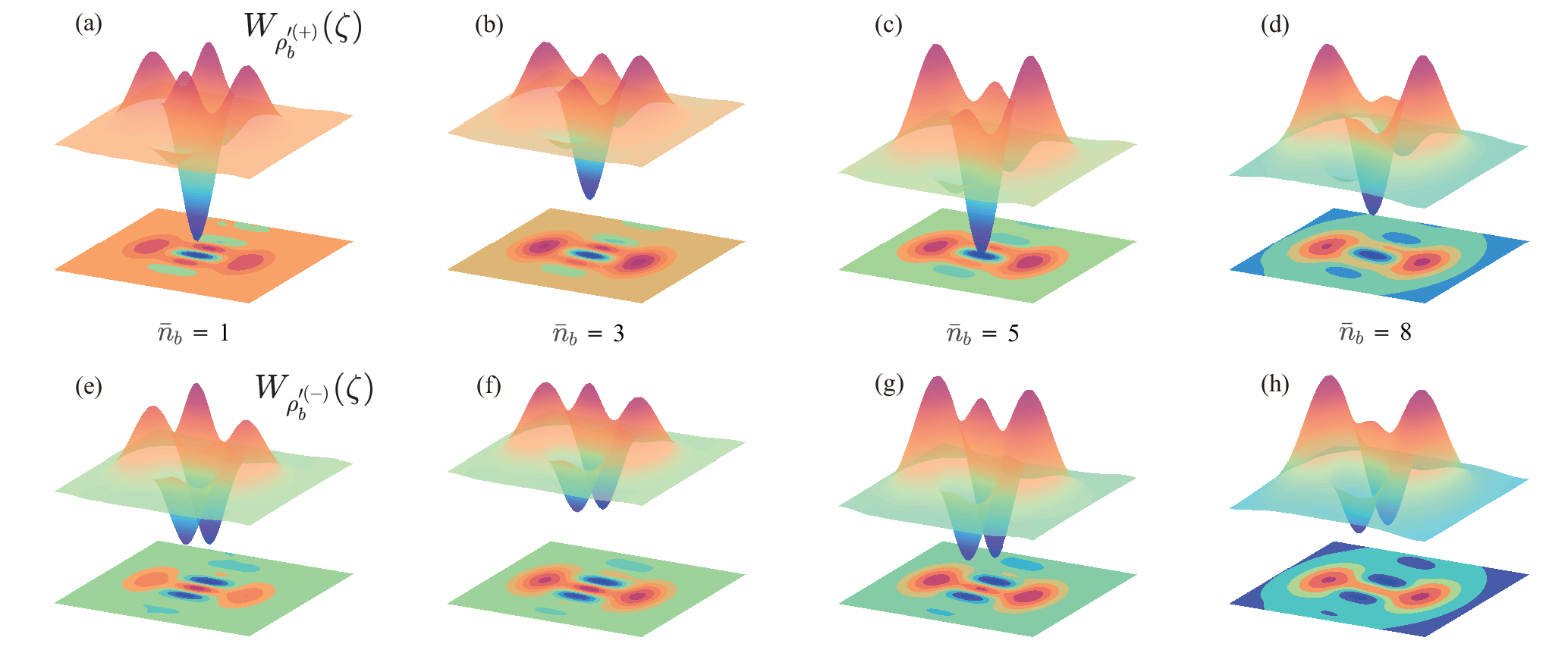}
\caption{Wigner functions $W_{\rho_{b}^{\prime(\pm)}}(\zeta)$ of the states (a)$-$(d) $\rho_{b}^{\prime(+)}(t_s)$ and (e)$-$(h) $\rho_{b}^{\prime(-)}(t_s)$ at different thermal excitation numbers $\bar{n}_{b}$: (a) and (e) $\bar{n}_{b}=1$, (b) and (f) $\bar{n}_{b}=3$, (c) and (g) $\bar{n}_{b}=5$, and (c) and (g) $\bar{n}_{b}=8$. The other parameters are $g/\Delta_{b}=0.001$, $\Delta_{c}/\Delta_{b}=20$, $\vert\xi_{\text{ss}}\vert=1700$, $\kappa_{o=a,b,c}/\Delta_{b}=0.01$, $\bar{n}_{a}=\bar{n}_{c}=0$, and $t_{s}=\pi/|\Delta_{b}|$.}
\label{Wigneropenvsnbfig}
\end{figure*}
%%%%%%%%%%%%%%%%%%%%%%%%%%%%%

Similar to the preceding discussion, by using Eq.~(\ref{the def of W finction}), we can obtain
the Winger functions $W_{\text{ext}}^{(\pm)}(\zeta)$ of the exact analytical cat states $|\psi_{\pm}(t)\rangle_{b}$ as
\begin{eqnarray}
W_{\text{ext}}^{(\pm )}(\zeta ) &=&\frac{2\left\vert \mathcal{K}
_{+}\right\vert ^{2}}{\pi }(e^{-2\left\vert \zeta \right\vert
^{2}}+e^{-2\left\vert \zeta -\beta _{2}(1)\right\vert ^{2}}  \notag \\
&&\pm 2\text{Re}[e^{-i\Theta _{\text{ext}}^{(1)}(t)}e^{-\left\vert \eta
_{2}(1)\right\vert ^{2}/2}e^{-i\text{Im}[-\zeta \beta _{2}^{\ast }\left(
1\right) ]}  \nonumber \\
&&\times e^{-2|\zeta |^{2}+(1/2)\zeta ^{\ast }\beta _{2}(1)+(3/2)\zeta \beta
_{2}^{\ast }(1)-(1/2)|\beta _{2}(1)|^{2}}]).
\end{eqnarray}
Using the relation given in Eq.~(\ref{relation for rorqudr}), the probability distributions $P_{\text{ext}}^{(\pm)}[X(\theta)]$ of the rotated quadrature operator $\hat{X}(\theta)$ for $|\psi_{\pm}(t)\rangle_{b}$ can be obtained as
\begin{eqnarray}
P_{\text{ext}}^{(\pm)}[X(\theta )]\!&=&\!|\,_{b}\langle X(\theta )|\psi _{\pm
}(t)\rangle _{b}|^{2}  \nonumber \\
\!&=&\!\mathcal{K}_{\pm}^{2}|_{b}\langle X(\theta )|0\rangle _{b}\pm e^{i\Theta _{
\text{ext}}^{(1)}(t)}e^{-\eta _{2}(1)|^{2}/2}\,_{b}\langle X(\theta )|\beta
_{2}\left( 1\right) \rangle _{b}\,|^{2}.  \nonumber \\
&&
\end{eqnarray}
Figures.~\ref{Wignerclosfig}(d)$-$\ref{Wignerclosfig}(f) plot the Wigner function $W_{\text{ext}}^{(+)}(\zeta)$ and the probability distribution $P_{\text{ext}}^{(+)}[X(\theta)]$ of the rotated quadrature operator $\hat{X}(\theta)$ at $|\xi_{\text{ss}}|=100$ and $2000$, which show good agreement with the results of the approximate analytical cat states $|\phi_{+}(t)\rangle_{b}$.

\subsection{Cat-state generation in the open-system case}
In order to confirm the scheme of the state generation in an ideal case, we consider the state generation based on the exact Hamiltonian $H_{\text{dis}}$ and including the system dissipation. Therefore, we need to numerically solve the quantum master equation and to measure modes $a$ and $c$ at time $t_{s}=\pi /|\Delta _{b}|$ in the states $|\pm \rangle _{a}|0\rangle _{c}$; then mode $b$ will collapse into two density matrices. To solve the evolution of the system, we express the density matrix in the Fock-state representation as
\begin{equation}
\label{MQ for cat-state}
\rho^{\prime}(t)=\sum_{m,j,s,n,k,r=0}^{\infty}\rho_{m,j,s,n,k,r}^{\prime}(t)|m\rangle_{a}|j\rangle_{b}|s\rangle _{c}\;_{a}\langle n|_{b}\langle k|_{c}\langle r|.
\end{equation}
The density matrix $\rho^{\prime}(t)$ can be obtained by solving the equations of motion for the density matrix elements. After the measurement, the reduced density matrices of mode $b$ become
\begin{eqnarray}
\rho_{b}^{\prime(\pm)}(t_{s})&=&\frac{\;_{a}\langle+|\;_{c}\langle 0|\rho^{\prime}|+\rangle_{a}|0\rangle _{c}}{\text{Tr}_{b}\left[\;_{a}\langle+|\;_{c}\langle 0|\rho^{\prime}|+\rangle_{a}|0\rangle_{c}\right]}
\nonumber \\
&=&\frac{1}{2P_{\pm }}\sum_{j,k=0}^{\infty }\mathcal{M}_{j,k}^{\pm
}|j\rangle_{b}\;_{b}\langle k|,
\end{eqnarray}
where we introduce the variables
\begin{equation}
\mathcal{M}_{j,k}^{\pm}=\rho_{0,j,0,0,k,0}^{\prime}\pm\rho_{0,j,0,1,k,0}^{\prime}\pm\rho_{1,j,0,0,k,0}^{\prime }+\rho_{1,j,0,1,k,0}^{\prime}
\end{equation}
and the measurement probabilities
\begin{equation}
P_{\pm }=\frac{1}{2}\sum_{j=0}^{\infty }\mathcal{M}_{j,j}^{\pm}.
\end{equation}
The fidelities between the generated cat states and the analytical target states are calculated by
\begin{equation}
F_{\pm}(t)=\,_{b}\langle \phi_{\pm }(t)|\rho_{b}^{\prime(\pm)}(t)|\phi_{\pm}(t)\rangle_{b}.
\end{equation}
Here we choose $|\xi_{\text{ss}}|=1700$ in our simulations so that $\vert\beta\vert_{\max}=2g_{0}/\vert\Delta _{b}\vert=3.4$; then the two states $|\beta(t)\rangle_{b}$ and $|0\rangle_{b}$ can be well distinguished in phase space $[_{b}\langle 0|\beta (t)\rangle _{b}=e^{-|\beta (t)|^{2}/2}\approx 10^{-3}$ at $|\beta|_{\text{max}}=3.4]$. In Fig.~\ref{fidpmopenfig}(a) we display the fidelities $F_{\pm}(t_{s})$ at the detection time $t_{s}=\pi/|\Delta_{b}|$ as functions of the decay rates $\kappa_{a}$ and $\kappa_{b}$, respectively. Here we can see that the influence of the decay of mode $b$ on the fidelities $F_{\pm}(t_{s})$ is more serious than that of mode $a$. In addition, we show $F_{\pm}(t_{s})$ versus the average thermal occupation $\bar{n}_{b}$ in Fig.~\ref{fidpmopenfig}(b). As expected, the fidelities are attenuated gradually with the increase of $\bar{n}_{b}$.

Using Eq.~(\ref{the def of W finction}), the Wigner functions for the states
$\rho_{b}^{\prime(\pm)}(t_{s})$ can be calculated as
\begin{equation}
\label{Wigner function}
W_{\rho _{b}^{\prime (\pm )}}(\zeta )=\frac{1}{\pi P_{\pm }}
\sum_{j,k,l=0}^{\infty }(-1)^{l}\mathcal{M}_{j,k\;b}^{\pm }\langle
l|D^{\dagger }(\zeta )|j\rangle _{b}\;_{b}\langle k|D(\zeta )|l\rangle _{b}.
\end{equation}
By calculating the matrix elements of the displacement operator in the number-state representation
\begin{equation}
_{b}\langle m|D\left( \zeta \right) |n\rangle _{b}=\left\{
\begin{array}{c}
\sqrt{\frac{m!}{n!}}e^{-|\zeta |^{2}/2}(-\zeta ^{\ast
})^{n-m}L_{m}^{n-m}(|\zeta |^{2}),\hspace{0.25cm}n>m \\
\sqrt{\frac{n!}{m!}}e^{-|\zeta |^{2}/2}(\zeta )^{m-n}L_{n}^{m-n}(|\zeta
|^{2}),\hspace{0.25cm}m>n,
\end{array}
\right.
\end{equation}
with $D\left(\zeta\right)=\exp(\zeta b^{\dagger}-\zeta^{\ast}b)$ and $L_{n}^{m}(x)$ the displacement operator and the associated Laguerre polynominals, repectively, then we can obtain the values of the Wigner functions $W_{\rho_{b}^{\prime(\pm)}}(\zeta )$.

To see the influence of the system dissipation on the Wigner functions, we show in Fig.~\ref{Wigneropenvsdecayfig} $W_{\rho_{b}^{\prime(\pm)}}(\zeta)$ of the generated cat states $\rho_{b}^{\prime(+)}(t_s)$ [Figs.~\ref{Wigneropenvsdecayfig}(a)$-$\ref{Wigneropenvsdecayfig}(d)] and $\rho_{b}^{\prime(-)}(t_s)$ [Figs.~\ref{Wigneropenvsdecayfig}(e)$-$\ref{Wigneropenvsdecayfig}(h)] in mode $b$ as a function of a complex variable $\zeta$ when the decay rate $\kappa_{a}$ takes different values. Obviously, the superposed coherent states with distinguishable superposition components and quantum interference pattern can be observed clearly from the Wigner functions. However, the increase of the decay rate $\kappa_{a}$ attenuates the interference pattern gradually (the region between the two peaks) and the main peak corresponding to the coherent component $|\beta\rangle $ is also reduced gradually.
%%%%%%%%%%%%%%%%%%%%%%%%%%%%%
\begin{figure}[tbp]
\center
\includegraphics[width=0.52\textwidth]{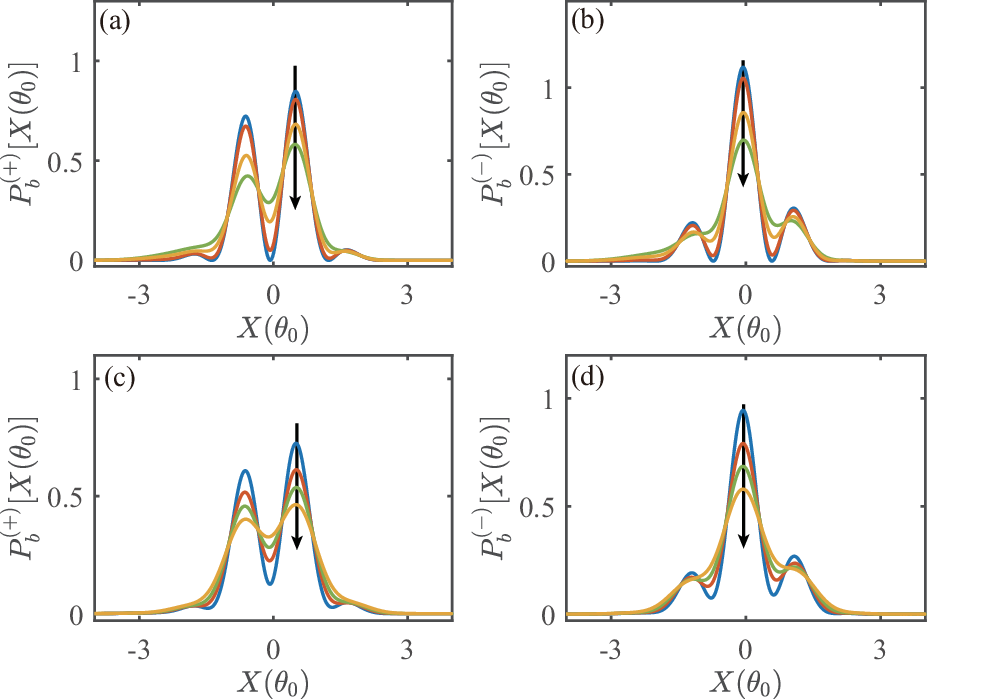}
\caption{Probability distributions $P_{b}^{\left(\pm\right)}[X(\theta_{0})]$ for the density matrices $\rho_{b}^{\prime(\pm)}(t_{s})$ as a function of $X(\theta_{0})$ in various cases: (a) and (b) $\bar{n}_{o=a,b,c}=0$ and $\kappa_{o=a,b,c}/\Delta_{b}=0$, 0.01, 0.05, and 0.1. (c) and (d) $\kappa_{o=a,b,c}/\Delta_{b}=0.01$ and $\bar{n}_{b}=1$, 3, 5, and 8. The other parameters are $g/\Delta_{b}=0.001$, $|\xi_{\text{ss}}|=1700$, $\Delta_{c}/\Delta_{b}=20$, $t_{s}=\pi/|\Delta_{b}|$, and $\theta_{0}=\arg[\beta(t_{s})]-\pi/2$.}
\label{probdisopenfig}
\end{figure}
%%%%%%%%%%%%%%%%%%%%%%%%%%%%%

As the effective frequency of mode $b$ is replaced by $\Delta_{b}$, acting as the frequency of a mechanical-like resonator, it makes sense to investigate the effect of the thermal occupation number $\bar{n}_{b}$ in mode $b$ on the generation of the cat states. In Fig.~\ref{Wigneropenvsnbfig} we display the Wigner functions  $W_{\rho_{b}^{\prime(\pm)}}(\zeta)$ of the generated cat states $\rho_{b}^{\prime(\pm)}(t_s)$ at different values of $\bar{n}_{b}$. We see that with the increase of $\bar{n}_{b}$, the quantum interference gradually decreases and even ultimately disappears.

Similarly to the closed-system case, from the probability distributions $P_{b}^{\left(\pm\right)}[X(\theta)]$, we can observe the influence of the dissipation on quantum interference. For the numerical cat states $\rho_{b}^{\prime(\pm)}(t_{s})$, the probability distributions can be obtained as
\begin{equation}
P_{b}^{(\pm )}[X(\theta )]=\frac{e^{-X^{2}(\theta )}}{2P_{\pm }}
\sum_{j,k=0}^{\infty }\frac{\mathcal{M}_{j,k}^{\pm }}{\sqrt{\pi 2^{j+k}j!k!}}
H_{j}[X(\theta )]H_{k}[X(\theta )]e^{i\theta (k-j)}.
\end{equation}

In Fig.~\ref{probdisopenfig} we show $P_{b}^{(\pm)}[X(\theta)]$ for the density matrices $\rho_{b}^{\prime(\pm)}(t_{s})$ as a function of $X(\theta_{0})$ when $\kappa_{o=a,b,c}/\Delta_{b}$ and $\bar{n}_{b}$ take different values. It can be seen that the oscillation amplitude of the probability distributions decrease gradually with the increase of the decay rates and thermal occupation number. This means that the dissipation of the system will wash out the quantum coherence in the cat states.

\section{Weak-to-strong transition of quantum measurement}\label{Weak-to-strong measurement}
The tunable optomechanical interaction is an ideal platform to show the  weak-to-strong transition in quantum measurement. Concretely, we choose modes $a$ and $b$ as the measured system and measuring pointer, respectively~\cite{Pepper2012}. Consider the preselection initial state $|i\rangle=|\psi_{\text{app}}(0)\rangle$; then the state of the system at time $t$ becomes $|\psi_{\text{app}}(t)\rangle$. In what follows, we perform a projective measurement by postselecting mode $a$ in the final state $|f\rangle=(\cos\vartheta|-\rangle_{a}+\sin\vartheta|+\rangle_{a})|0\rangle_{c}$, where $\vartheta$ is the postselection angle. To obtain the measurement information of the system, we measure the operator $a^{\dagger}a$ in both the weak- and strong-coupling cases. According to the definition of a weak value~\cite{Aharonov1998,Kofman2012,Dressel2014}, we have $\langle a^{\dagger }a\rangle _{\text{W}}=\langle f|a^{\dagger }a|i\rangle/\langle f|i\rangle =(1-\cot\vartheta)/2$ in the weak-measurement regime. However, in the strong-measurement regime, we gain the expectation value as $\langle a^{\dagger }a\rangle _{\text{S}}=\langle f|a^{\dagger }a|f\rangle/\langle f|f\rangle =[1-\sin (2\vartheta)]/2$.
Performing the projective measurement of state $|\psi_{\text{app}}(t)\rangle$ in the final state $|f\rangle$, the normalized final state of the pointer becomes
\begin{equation}
\label{cat state}
|\mathrm{cat}_{\vartheta}\rangle_{b}=\frac{\sin(\vartheta+\frac{\pi }{4}) |0\rangle _{b}-\cos(\vartheta +\frac{\pi}{4})
e^{i\varphi(t)}|\beta(t)\rangle_{b}}{\sqrt{1-e^{-\left\vert\beta(t)\right\vert^{2}/2}\cos[\varphi(t)]\cos(2\vartheta)}},
\end{equation}
where the factor $_{b}\langle 0|\beta(t)\rangle_{b}=e^{-\left\vert\beta(t)\right\vert^{2}/2}$ determines the transition from weak to strong measurements.
%%%%%%%%%%%%%%%%%%%%%%%%%%%%%
\begin{figure}[tbp]
\center\includegraphics[width=0.48\textwidth]{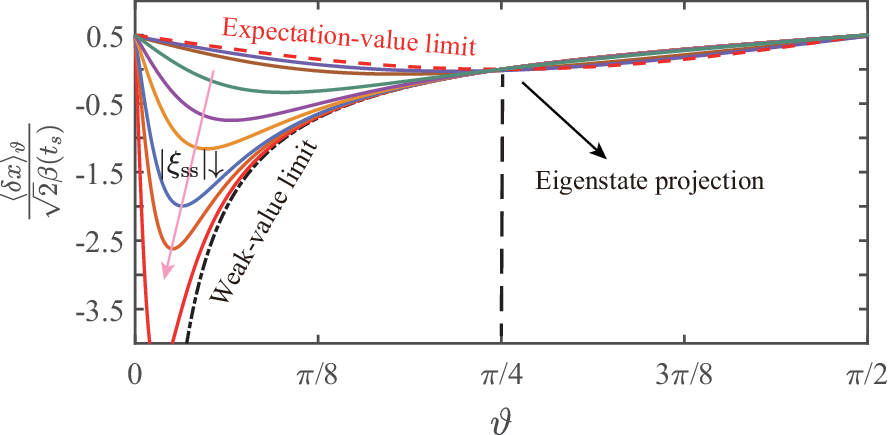}
\caption{Pointer shift $\langle\delta x\rangle_{\vartheta}/\sqrt{2}\beta(t_{s})$ as a function of the postselection angle $\vartheta$ at $|\xi_{\text{ss}}|=50$, 80, 100, 150, 200, 300, 500, and 600. The other parameters are $g/\Delta_{b}=0.001$, $t_{s}=\pi/|\Delta_{b}|$, and $\omega_{a}/\Delta_{b}=2n$ with $n$ a non-negative integer.}
\label{pointershitfig}
\end{figure}
%%%%%%%%%%%%%%%%%%%%%%%%%%%%%

%%%%%%%%%%%%%%%%%%%%%%%%%%%%%
\begin{figure*}[tbp]
\center
\includegraphics[width=0.8 \textwidth]{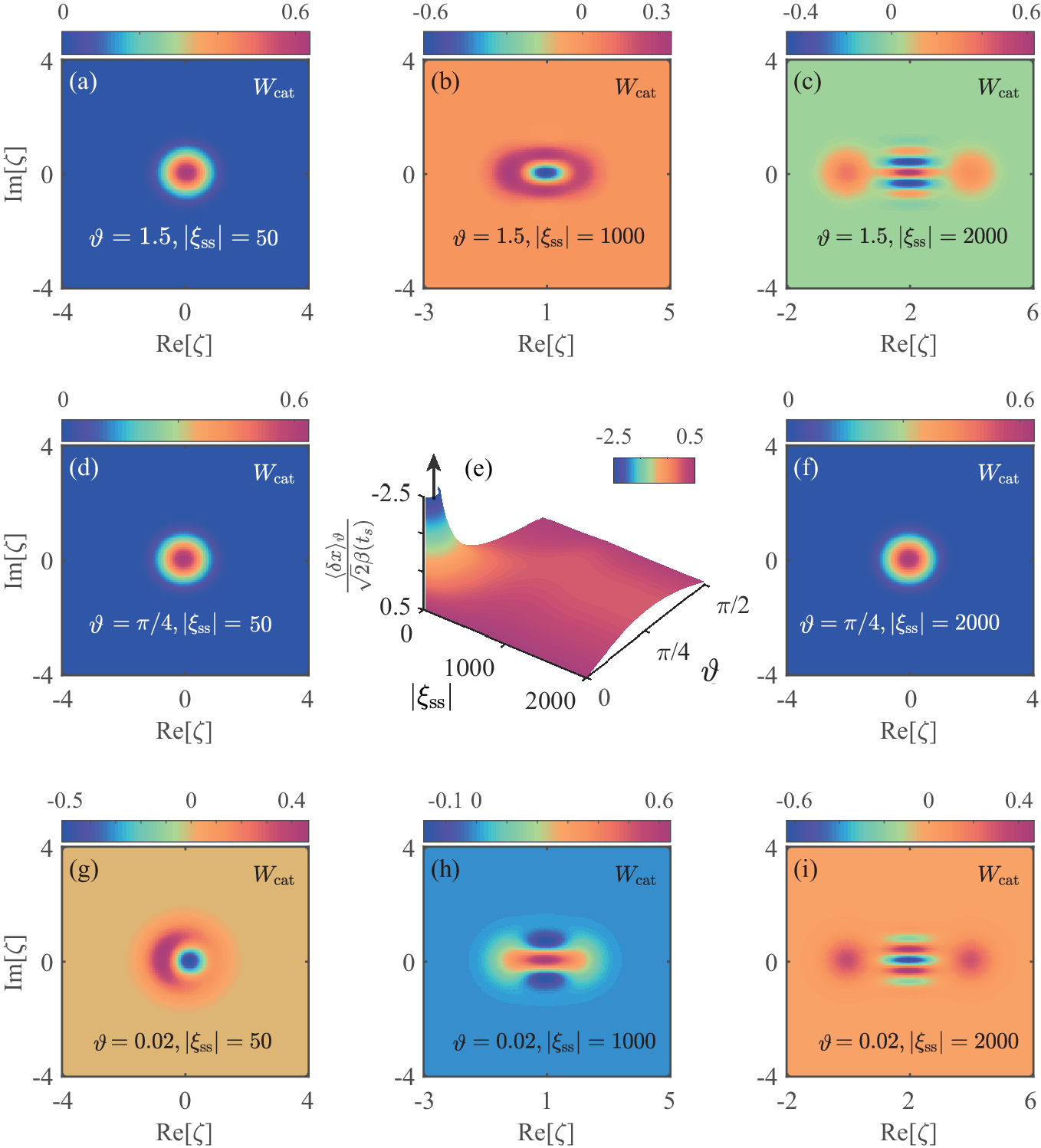}
\caption{Measurement regimes of $\langle\delta x\rangle_{\vartheta}/\sqrt{2}\beta(t_{s})$ in the full parameter space $(\vert\xi_{\text{ss}}\vert, \vartheta)$. The Wigner functions of the pointer states $|\text{cat}_{\vartheta}\rangle_{b}$ are plotted in various regimes: (a), (d), and (g) the weak-measurement regime with $\vert\xi_{\text{ss}}\vert=50$; (b) and (h) the intermediated-measurement regime with $\vert\xi_{\text{ss}}\vert=1000 $; and (c), (f), and (i) the strong-measurement regime with $\vert\xi_{\text{ss}}\vert=2000$. The postselection angles $\vartheta=0.02$, $\pi/4$, and $1.5$ correspond to nearly orthogonal, eigenstate projection, and nearly parallel pre- and post-selected states, respectively. (e) Cat state shift $\langle\delta x\rangle_{\vartheta}/\sqrt{2}\beta(t_{s})$ as a function of $\vert\xi_{\text{ss}}\vert$ and $\vartheta$. The other parameters are $g/\Delta_{b}=0.001$, $t_{s}=\pi/|\Delta_{b}|$, and $\omega_{a}/\Delta_{b}=2n$, with $n$ a non-negative integer.}
\label{pointerWignerfig}
\end{figure*}
%%%%%%%%%%%%%%%%%%%%%%%%%%%%%
The above two cases of measurement outcomes can be unified by the pointer shift relative to that of the vacuum state $|0\rangle_{b}$, which is defined as $\langle\delta x\rangle_{\vartheta}\equiv\;_{b}\langle \mathrm{cat}_{\vartheta}|\hat{x}|\mathrm{cat}_{\vartheta}\rangle _{b}/_{b}\langle\mathrm{cat}_{\vartheta}|\mathrm{cat}_{\vartheta}\rangle _{b}$, with the dimensionless position operator defined as $\hat{x}=(a^{\dagger}+a)/\sqrt{2}$. Substitution of Eq.~(\ref{cat state}) into $\langle\delta x\rangle_{\vartheta}$ yields
\begin{equation}
\label{cat displacement}
\langle \delta x\rangle _{\vartheta }=\frac{\beta (t_{s})}{\sqrt{2}}\left(\! 1\!-\!
\frac{\sin (2\vartheta )}{1-e^{-\left\vert \beta (t_{s})\right\vert
^{2}/2}\cos[\varphi (t_{s})]\cos (2\vartheta )}\!\right)\!.
\end{equation}
Here we choose the detection time $t_{s}=\pi/|\Delta_{b}|$ and the phase angle $\theta_{c}=0$. Then we have $\beta(t_{s})=\beta^{\ast}( t_{s})=2g_{0}/|\Delta_{b}|$ with $g_{0}=g\vert\xi_{\text{ss}}\vert$. In the weak-coupling regime $\beta(t_{s})\ll 1$, the factor $\cos[\varphi(t)]$ in Eq.~(\ref{cat displacement}) approaches 1 at proper parameter conditions. We then obtain $\left. \left\langle\delta x\right\rangle_{\vartheta}\right\vert_{\beta(t_{s})\rightarrow 0}=\sqrt{2}\beta(t_{s})\langle a^{\dagger}a\rangle_{\text{W}}$, which is consistent with the weak value of the measured photon number operator. In the strong-coupling regime $\beta(t_{s})\gg 1$, we find $\left. \left\langle \delta x\right\rangle_{\vartheta}\right\vert_{\beta(t_{s})\rightarrow \infty}=\sqrt{2}\beta(t_{s})\langle a^{\dagger}a\rangle_{\text{S}}$, corresponding to the expectation value. In Fig.~\ref{pointershitfig} we display the pointer shift $\left\langle\delta x\right\rangle_{\vartheta}/\sqrt{2}\beta(t_{s})$ as a function of the postselection angle $\vartheta$ at different values of $\vert\xi_{\text{ss}}\vert$. It can be found that when $\vert\xi_{\text{ss}}\vert$ is small, we can appropriately choose the postselection angle such that the pre- and postselected states are nearly orthogonal. Then there exists a surprising weak-value amplification phenomenon~\cite{Aharonov1998,Kofman2012,Dressel2014}. In particular, the weak value is equal to the expectation value at $\vartheta=\pi/4$, indicating an eigenstate projection. Moreover, by tuning the enhanced optomechanical coupling strength $g_{0}$, the pointer shift $\langle\delta x\rangle_{\vartheta}$ will show the transition of the measurement outcome from a weak value to an expectation value~\cite{Pan2020}.

To clearly see the weak-to-strong transition of quantum measurement, we plot the Wigner functions of the pointer state in the full parameter space $(\vert\xi_{\text{ss}}\vert, \vartheta)$ for the measurement regime in Fig.~\ref{pointerWignerfig}. Figures~\ref{pointerWignerfig}(a),~\ref{pointerWignerfig}(d), and~\ref{pointerWignerfig}(g) correspond to the weak-measurement regime with $\vert\xi_{\text{ss}}\vert=50$ and post-selection angles $\vartheta=0.02$, $\pi/4$, and $1.5$, respectively. The three angles represent, respectively, nearly orthogonal, eigenstate projection, and nearly parallel pre- and postselected states. As shown in Fig.~\ref{pointerWignerfig}(g), for the postselection angle $\vartheta=0.02$, the postselection state is nearly orthogonal to the preselected state Then a remarkable weak-value amplification can be observed even in the weak-measurement coupling regime with $\vert\xi_{\text{ss}}\vert=50$. Figures~\ref{pointerWignerfig}(b) and~\ref{pointerWignerfig}(h) represent the intermediate-measurement regime at post-selection angles $\vartheta=1.5$ and $0.02$, from which we can still see a distinct overlap between the two superposed coherent states. In Figs.~\ref{pointerWignerfig}(c) and~\ref{pointerWignerfig}(i), however, the two coherent states can have a negligible overlap as the increase of the steady-state displacement amplitude $\vert\xi_{\text{ss}}\vert$. Therefore, the two well-separated coherent states can be observed in phase space. In Figs.~\ref{pointerWignerfig}(d) and~\ref{pointerWignerfig}(f) we see there is only one peak located at the origin of the phase space. This is because $\vartheta=\pi/4$ corresponds to the eigenstate projection, which can cause the superposition coefficient of the coherent state $|\beta(t_{s})\rangle_{b}$ in the cat state $|\text{cat}_{\vartheta}\rangle_{b}$ to be 0. Then the cat state reduces to the ground state $|0\rangle_{b}$ of mode $b$. Hence, we cannot observe the displacement of the cat state for all coupling strengths.

\section{Photon blockade effect in mode $a$}\label{photon blockade effect}
In this section we study the photon blockade effect in the simulated ultrastrong optomechanical system. To observe the photon blockade effect, we introduce a monochromatic weak-driving field to mode $a$. Then in a rotating frame with respect to $H_{0}=\omega_{d}a^{\dagger}a+\omega_{L}b^{\dagger}b+\omega_{L}c^{\dagger}c$, the Hamiltonian of the total system reads
\begin{eqnarray}
H_{L}^{\prime}&=&\Delta_{a}a^{\dagger}a+\Delta_{b}b^{\dagger}b+\Delta_{c}c^{\dagger}c+ga^{\dagger}a(b^{\dagger }c+c^{\dagger }b)\nonumber\\
&&+\Omega_{a}a^{\dagger}+\Omega_{a}^{\ast}a+\Omega_{c}c^{\dagger}+\Omega_{c}^{\ast}c,
\end{eqnarray}
where we introduce the driving detunings $\Delta_{a}=\omega_{a}-\omega_{d}$, $\Delta_{b}=\omega_{b}-\omega_{L}$, and $\Delta_{c}=\omega_{c}-\omega_{L}$. The driving amplitude and frequency of mode $a$ ($c$) are $\Omega_{a}$ ($\Omega_{c}$) and $\omega_{d}$ ($\omega_{L}$), respectively. Note that the driving field on mode $a$ is weak, i.e., $\Omega_{a}/\kappa_{a}\ll 1$, so we treat the driving of mode $a$ as a perturbation~\cite{Liao2013}. It should be pointed out that the driving on mode $c$\ is strong while the driving on mode $a$ is weak; therefore, the displacement transformation is only performed on mode $c$ as $\rho^{\prime}=D_{c}^{\dagger}(\zeta)\rho D_{c}(\zeta)$. In the open-system case, we obtain the quantum master equation in the displacement representation as

\begin{equation}
\label{QMEQ for PB}
\dot{\rho}^{\prime}=i[\rho^{\prime},H^{\prime}_{\text{dis}}]+\sum_{o=a,b,c}\{\kappa_{o}(\bar{n}_{o}+1)
\mathcal{D}[o]\rho^{\prime}+\kappa_{o}\bar{n}_{o}\mathcal{D}[o^{\dagger}]\rho^{\prime}\},
\end{equation}
with the Hamiltonian in the displacement representation
\begin{eqnarray}
H_{\text{dis}}^{\prime}&=&\Delta_{a}a^{\dagger}a+\Delta_{b}b^{\dagger}b+\Delta_{c}c^{\dagger}c+ga^{\dagger }a(b^{\dagger}c+c^{\dagger}b)\nonumber\\
&&-g_{0}a^{\dagger}a(b^{\dagger}e^{i\theta_{c}}+be^{-i\theta_{c}})+\Omega_{a}a^{\dagger}+\Omega_{a}^{\ast}a,
\end{eqnarray}
where we introduce the single-photon optomechanical coupling strength $g_{0}=g|\xi_{\text{ss}}|$. To analyze the photon blockade effect in mode $a$, we first analytically calculate the equal-time second-order correlation function $g^{(2)}(0)$ of mode $a$.

To study the photon blockade effect, we first diagonalize the undriven Hamiltonian
\begin{eqnarray}
\label{undrivenHamiltonian}
H_{\text{dis}}^{\prime\prime}&=&\Delta_{a}a^{\dagger}a+\Delta_{b}b^{\dagger}b+\Delta_{c}c^{\dagger }c+ga^{\dagger}a(b^{\dagger}c+c^{\dagger}b)\nonumber\\
&&-g_{0}a^{\dagger}a(b^{\dagger}+b),
\end{eqnarray}
where we have selected the phase angle $\theta_{c}=0$. To this end, we introduce the displacement operators $D_{b}(\beta)=e^{\beta(b^{\dagger}-b)}$ and $D_{c}(\eta)=e^{\eta(c^{\dagger}-c)}$ and  the transformation operator $T=e^{\lambda(b^{\dagger}c-c^{\dagger}b)}$. Using Eqs.~(\ref{lambda})$-$(\ref{chi}), the diagonalized Hamiltonian can be obtained as
\begin{eqnarray}
\tilde{H}_{\text{dis}}^{\prime\prime}&=&D_{c}^{\dagger}(\eta)D_{b}^{\dagger}(\beta)T^{\dagger }H_{\text{dis}}^{\prime\prime}TD_{b}(\beta)D_{c}(\eta)\nonumber\\
&=&\Delta_{a}a^{\dagger}a+\chi_{b}b^{\dagger}b+\chi_{c}c^{\dagger}c\nonumber\\
&&-\frac{g_{0}^{2}\cos^{2}\lambda}{\chi_{b}}a^{\dagger}aa^{\dagger}a-
\frac{g_{0}^{2}\sin^{2}\lambda}{\chi_{c}}a^{\dagger}aa^{\dagger}a.
\end{eqnarray}
Then the eigensystem of the undriven Hamiltonian $H_{\text{dis}}^{\prime\prime}$ is given by
\begin{equation}
\label{eigensystem of unHdr}
H_{\text{dis}}^{\prime\prime}T\vert m\rangle_{a}\vert\tilde{j}(m)\rangle_{b}\vert\tilde{s}(m)\rangle_{c}=E_{m,j,s}T\vert m\rangle_{a}\vert \tilde{j}(m)\rangle_{b}\vert\tilde{s}(m)\rangle_{c},
\end{equation}
where the photon-number-dependent Fock states of modes $b$ and $c$ are defined by
\begin{subequations}
\label{Fock state}
\begin{align}
|\tilde{j}(m)\rangle_{b}&=\exp[{\beta(m)(b^{\dagger}-b)}]|j\rangle_{b},\\
|\tilde{s}(m)\rangle_{c}&=\exp[{\eta(m)(c^{\dagger}-c)}]|s\rangle_{c}.
\end{align}
\end{subequations}
The eigenvalues in Eq.~(\ref{eigensystem of unHdr}) are defined by
\begin{eqnarray}
\label{eigenenergy spectrum}
E_{m,j,s}&=&\Delta_{a}m+\chi_{b}(m)j+\chi_{c}(m)s\nonumber\\
&&-\frac{g_{0}^{2}\cos^{2}[\lambda(m)]}{\chi_{b}(m)}m^{2}-\frac{g_{0}^{2}\sin^{2}[\lambda(m)]}{\chi_{c}(m)}m^{2},
\end{eqnarray}
which show the photonic nonlinearity in the eigenenergy spectrum. This photonic nonlinearity is the physical origin of photon blockade effect.

To include the influence of the system dissipation on photon blockade, we phenomenologically introduce a non-Hermitian term to the Hamiltonian~(\ref{undrivenHamiltonian}),
\begin{equation}
H_{\text{eff}}=H_{\text{dis}}^{\prime}-i\frac{\kappa_{a}}{2}a^{\dagger}a,
\end{equation}
where $\kappa_{a}$ is the decay rate of mode $a$. In the weak-driving case, i.e., $\Omega_{a}\ll \kappa_{a}$, we can restrict the system to the few-photon subspace spanning these basis states $\{{\vert 0\rangle_{a},\vert 1\rangle_{a},\vert 2\rangle_{a}}\}$. A general state of the system in this subspace can then be written as
\begin{eqnarray}
|\psi (t)\rangle&=&T|\varphi(t)\rangle\nonumber\\
&=&\sum_{f,h=0}^{\infty}C_{0,f,h}(t)|0\rangle_{a}|f\rangle_{b}|h\rangle_{c}\nonumber\\
&&+\sum_{f,h=0}^{\infty}C_{1,f,h}(t)e^{\lambda(1)(b^{\dagger}c-c^{\dagger}b)}|1\rangle_{a}|\tilde{f}(1)\rangle _{b}|\tilde{h}(1)\rangle_{c}\nonumber\\
&&+\sum_{f,h=0}^{\infty}C_{2,f,h}(t)e^{\lambda(2)(b^{\dagger}c-c^{\dagger}b)}|2\rangle_{a}|\tilde{f}(2)\rangle _{b}|\tilde{h}(2)\rangle_{c},
\end{eqnarray}
where $C_{0,f,h}(t)$, $C_{1,f,h}(t)$, and $C_{2,f,h}(t)$ are the probability amplitudes corresponding to the basis states $\vert 0\rangle_{a}\vert f\rangle_{b}\vert h\rangle_{c}$, $e^{\lambda(1)(b^{\dagger}c-c^{\dagger}b)}\vert 1\rangle_{a}\vert\tilde{f}(1) \rangle _{b}\vert\tilde{h}(1)\rangle_{c}$, and $e^{\lambda(2)(b^{\dagger}c-c^{\dagger}b)}\vert 2\rangle_{a}\vert \tilde{f}(2)\rangle_{b}\vert \tilde{h}(2)\rangle_{c}$, respectively. In terms of the Schr\"{o}dinger equation $i\vert\dot{\psi}(t)\rangle=H_{\text{eff}}\vert\psi(t)\rangle$, the equations of motion for these probability amplitudes can be obtained by
\begin{subequations}
\label{amplitudeEq}
\begin{eqnarray}
\dot{C}_{0,m,n}&=&-iE_{0,m,n}C_{0,m,n}-i\Omega_{a}\sum_{f,h=0}^{\infty}B_{1}C_{1,f,h},\\
\dot{C}_{1,m,n}&=&-i(E_{1,m,n}-i\kappa_{a}/2)C_{1,m,n}-i\Omega_{a}\sum_{f,h=0}^{\infty}B_{2}C_{0,f,h}\nonumber\\
&-&i\sqrt{2}\Omega_{a}\sum_{f,h=0}^{\infty}B_{3}C_{2,f,h},\\
\dot{C}_{2,m,n}&=&-i(E_{2,m,n}-i\kappa_{a})C_{2,m,n}-i\sqrt{2}\Omega_{a}\sum_{f,h=0}^{\infty }B_{4}C_{1,f,h},\nonumber\\
\end{eqnarray}
\end{subequations}
where $B_{1}$, $B_{2}$, $B_{3}$, and $B_{4}$ are defined by
\begin{subequations}
\begin{align}
B_{1}& =\ _{b}\!\langle m|_{c}\langle n|e^{\lambda (1)(b^{\dagger
}c-c^{\dagger }b)}|\tilde{f}(1)\rangle _{b}|\tilde{h}(1)\rangle _{c}, \\
B_{2}& =\ _{b}\!\langle \tilde{m}(1)|_{c}\langle \tilde{n}(1)|e^{-\lambda
(1)(b^{\dagger }c-c^{\dagger }b)}|f\rangle _{b}|h\rangle _{c}, \\
B_{3}& =\ _{b}\!\langle \tilde{m}(1)|_{c}\langle \tilde{n}(1)|e^{[\lambda
(2)-\lambda (1)](b^{\dagger }c-c^{\dagger }b)}|\tilde{f}(2)\rangle _{b}|
\tilde{h}(2)\rangle _{c}, \\
B_{4}& =\ _{b}\!\langle \tilde{m}(2)|_{c}\langle \tilde{n}(2)|e^{-[\lambda
(2)-\lambda (1)](b^{\dagger }c-c^{\dagger }b)}|\tilde{f}(1)\rangle _{b}|
\tilde{h}(1)\rangle_{c}.
\end{align}
\end{subequations}
%%%%%%%%%%%%%%%%%%%%%%%%%%%%%
\begin{figure}[tbp]
\center
\includegraphics[width=0.48\textwidth]{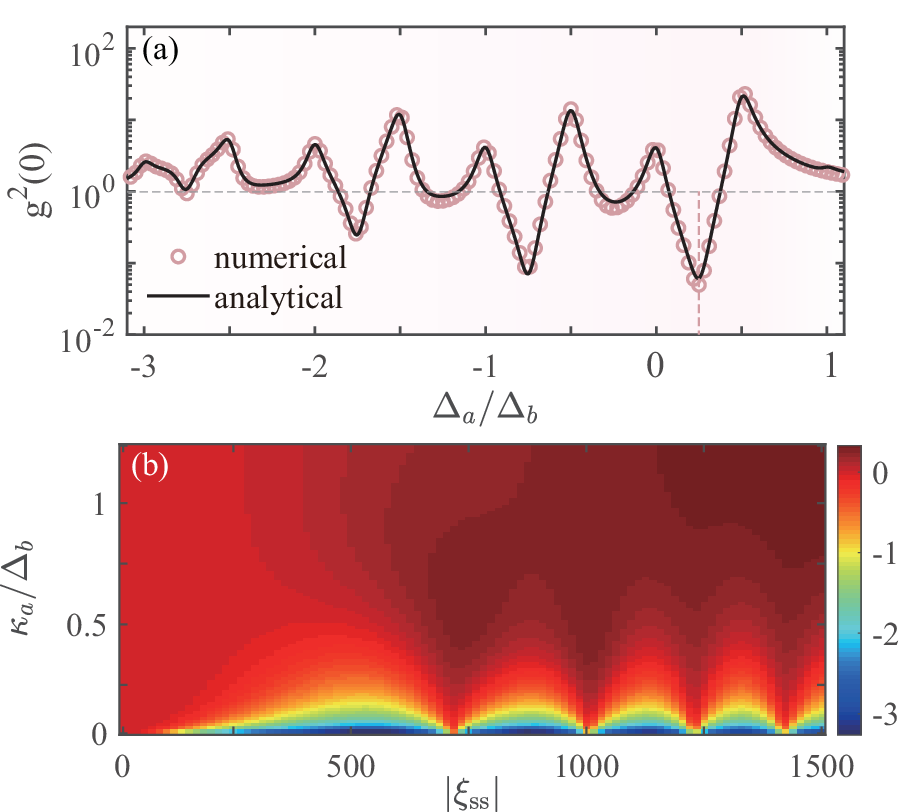}
\caption{(a) Plot of the equal-time second-order correlation function $g^{(2)}(0)$ as a function of the driving detuning $\Delta_{a}/\Delta_{b}$ for $|\xi_{\text{ss}}|=500$ and $\kappa_{a}/\Delta_{b}=0.1$. The black solid line and the circles represent the numerical and analytical results, respectively. (b) Plot of $\log_{10}g^{(2)}(0)$ as a function of the decay rate $\kappa_{a}/\Delta_{b}$ and the steady-state displacement $\vert\xi_{\text{ss}}\vert$ at $\kappa_{b}/\Delta_{b}=\kappa_{c}/\Delta_{b}=0.001$ and $\bar{n}_{o=a,b,c}=0$. Here, the single-photon resonance condition $\Delta_{a}=g_{0}^2/\Delta_{b}$ is considered. The other parameters are $g/\Delta_{b}=0.001$, $\Delta_{c}/\Delta_{b}=20$, and $\Omega_{a}/\kappa_{a}=0.1$.}
\label{anaandnumg2fig}
\end{figure}
%%%%%%%%%%%%%%%%%%%%%%%%%%%%%

In the weak-driving case, we use the perturbation method to approximately solve Eq.~(\ref{amplitudeEq}) by discarding the
higher-order terms in the equations of motion for the lower-order variables. Considering the initial state $\vert 0\rangle_{a}\vert 0\rangle_{b}\vert 0\rangle_{c}$ of the system, i.e., $C_{0,j,s}(0)=\delta_{j,0}\delta_{s,0}$, the long-time solutions of Eq.~(\ref{amplitudeEq}) are approximately obtained as
\begin{subequations}
\begin{align}
C_{0,m,n}(\infty)&=\delta_{m,0}\delta_{n,0},\\
C_{1,m,n}(\infty)&=\frac{-\Omega_{a}D_{1}}{E_{1,m,n}-i\kappa_{a}/2}, \\
C_{2,m,n}(\infty)&=\sum_{f,h=0}^{\infty}\frac{\;\sqrt{2}\Omega
_{a}^{2}D_{2}D_{3}}{(E_{1,f,h}-i\kappa_{a}/2)(E_{2,m,n}-i\kappa_{a})},
\end{align}
\end{subequations}
where $D_{1}$, $D_{2}$, and $D_{3}$ are defined by
\begin{subequations}
\begin{align}
D_{1}&=\ _{b}\!\langle\tilde{m}(1)|_{c}\langle\tilde{n}(1)|e^{-\lambda
(1)(b^{\dagger}c-c^{\dagger}b)}|0\rangle_{b}|0\rangle_{c},\\
D_{2}&=\ _{b}\!\langle\tilde{f}(1)|_{c}\langle\tilde{h}(1)|e^{-\lambda
(1)(b^{\dagger}c-c^{\dagger}b)}|0\rangle_{b}|0\rangle_{c},\\
D_{3}&=\ _{b}\!\langle\tilde{m}(2)|_{c}\langle\tilde{n}(2)|e^{-[\lambda
(2)-\lambda(1)](b^{\dagger}c-c^{\dagger}b)}|\tilde{f}(1)\rangle_{b}|
\tilde{h}(1)\rangle_{c}.
\end{align}
\end{subequations}

In the long-time limit, the single- and two-photon probabilities can be obtained as
\begin{subequations}
\label{amplitudeP}
\begin{align}
P_{1} =&\frac{1}{\mathcal{N}}\sum_{m,n=0}^{\infty }\left\vert \frac{\Omega
_{a}D_{1}}{(E_{1,m,n}-i\kappa _{a}/2)}\right\vert ^{2},\\
P_{2} =&\frac{1}{\mathcal{N}}\sum_{m,n=0}^{\infty }\left\vert
\sum_{f,h=0}^{\infty }\frac{\sqrt{2}\Omega _{a}^{2}D_{2}D_{3}}{
(E_{1,f,h}-i\kappa _{a}/2)(E_{2,m,n}-i\kappa _{a})}\right\vert^{2},
\end{align}
\end{subequations}
with the normalization constant
\begin{eqnarray}
\mathcal{N} &=&1+\sum_{m,n=0}^{\infty }\left\vert \frac{\Omega _{a}D_{1}}{
(E_{1,m,n}-i\kappa _{a}/2)}\right\vert ^{2} \nonumber \\
&&+\sum_{m,n=0}^{\infty }\left\vert \sum_{f,h=0}^{\infty }\frac{\sqrt{2}
\Omega _{a}^{2}D_{2}D_{3}}{(E_{1,f,h}-i\kappa _{a}/2)(E_{2,m,n}-i\kappa_{a})}\right\vert ^{2}.
\end{eqnarray}

In the weak-driving case, the equal-time second-order correlation function $g^{(2)}(0)$ can be obtained as
\begin{equation}
g^{(2)}(0)\equiv\frac{\langle a^{\dagger}a^{\dagger}aa\rangle}{\langle a^{\dagger}a\rangle^{2}}=\frac{2P_{2}}{
(P_{1}+2P_{2})^{2}}\approx\frac{2P_{2}}{P_{1}^{2}},\label{anag2}
\end{equation}
which is determined by the single- and two-photon probabilities. To calculate the probabilities $P_{1}$ and $P_{2}$, we need to calculate the matrix elements $_{b}\langle \tilde{m}(1)|_{c}\langle \tilde{n}(1)|e^{-\lambda(1)(b^{\dagger}c-c^{\dagger}b)}|0\rangle_{b}|0\rangle_{c}$ and $_{b}\langle \tilde{m}(2)|_{c}\langle \tilde{n}(2)|e^{-[\lambda(2)-\lambda(1)](b^{\dagger}c-c^{\dagger}b)}|\tilde{f}(1)\rangle_{b}|\tilde{h}( 1)\rangle_{c}$. Using the relations given in Eq.~(\ref{Fock state}), the first term can still be calculated with the Laguerre polynomial $L_{n}^{m}(x)$ because it can be further reduced to
\begin{eqnarray}
&&\;_{b}\langle \tilde{m}(1)|\;_{c}\langle \tilde{n}(1)|e^{-\lambda
(1)(b^{\dagger }c-c^{\dagger }b)}|0\rangle _{b}|0\rangle _{c}  \nonumber \\
&=&\;_{b}\langle m|e^{-\beta (1)(b^{\dagger }-b)}|0\rangle _{b}\;_{c}\langle
n|e^{-\eta (1)(c^{\dagger }-c)}|0\rangle _{c},
\end{eqnarray}
where we have used the relation $e^{-\lambda (1)(b^{\dagger }c-c^{\dagger }b)}|0\rangle _{b}|0\rangle_{c}=|0\rangle _{b}|0\rangle _{c}$. However, the second term
\begin{eqnarray}
\label{matrix elements2}
&&\;_{b}\langle\tilde{m}(2)|_{c}\langle\tilde{n}(2)|e^{-[\lambda(2)-\lambda (1)](b^{\dagger }c-c^{\dagger }b)}|\tilde{f}(1)\rangle _{b}|\tilde{h}(1)\rangle_{c}\nonumber\\
&=&\;_{b}\langle m|_{c}\langle n|e^{-\beta (2)(b^{\dagger }-b)}e^{-\eta
(2)(c^{\dagger }-c)}e^{-[\lambda (2)-\lambda (1)](b^{\dagger }c-c^{\dagger}b)}\nonumber\\
&\times&e^{\beta (1)(b^{\dagger }-b)}e^{\eta (1)(c^{\dagger }-c)}|f\rangle_{b}|h\rangle_{c}
\end{eqnarray}
needs to be calculated numerically. According to Eq.~(\ref{lambdambetaandetamchim}), the expressions of $\beta(1)$, $\eta(1)$, $\beta(2)$, and $\eta(2)$ are given by
\begin{subequations}
\label{beta1}
\begin{align}
\beta(n)=&\frac{ng_{0}\cos[\lambda(n)]}{\Delta_{b}\cos^{2}[\lambda(n)]+\Delta_{c}\sin^{2}[\lambda(n)]-ng\sin [2\lambda(n)]},\\
\eta(n)=&\frac{ng_{0}\sin[\lambda(n)]}{\Delta_{b}\sin^{2}[\lambda(n)]+\Delta_{c}\cos^{2}[\lambda(n)]+ng\sin [2\lambda (n)]},
\end{align}
\end{subequations}
with $\lambda (n)=\arctan [2ng/(\Delta _{c}-\Delta _{b})]/2$ for $n=1,2$. By substituting Eq.~(\ref{amplitudeP}) into Eq.~(\ref{anag2}), we can obtain the result of the equal-time second-order correlation function $g^{(2)}(0)$.

To go beyond the analytical result obtained with the perturbation method, we also calculate the second-order correlation function $g^{(2)}(0)$ by numerically solving the steady state of the quantum master equation~(\ref{QMEQ for PB}). We define the steady-state density matrix of the system as
\begin{equation}
\tilde{\rho}^{\text{ss}}=\!\!\sum_{m,j,s,n,k,r=0}^{\infty}\tilde{\rho}_{m,j,s,n,k,r}^{\text{ss}}|m\rangle_{a}|j\rangle_{b}|s\rangle_{c}\;_{a}\langle n|_{b}\langle k|_{c}\langle r|.
\end{equation}
Then the equal-time second-order correlation function $g^{2}(0)$ can be obtained by
\begin{equation}
\label{numericalg2}
g^{(2)}(0)=\frac{\langle a^{\dagger}a^{\dagger}aa\rangle_{\text{ss}}}{\langle a^{\dagger}a\rangle_{\text{ss}}^{2}}=\frac{\text{Tr}
[a^{\dagger}a^{\dagger}aa\tilde{\rho}^{\text{ss}}]}{(\text{Tr}[a^{\dagger}a\tilde{\rho}^{\text{ss}}])^{2}}.
\end{equation}
%%%%%%%%%%%%%%%%%%%%%%%%%%%%%
\begin{figure}[tbp]
\center
\includegraphics[width=0.48\textwidth]{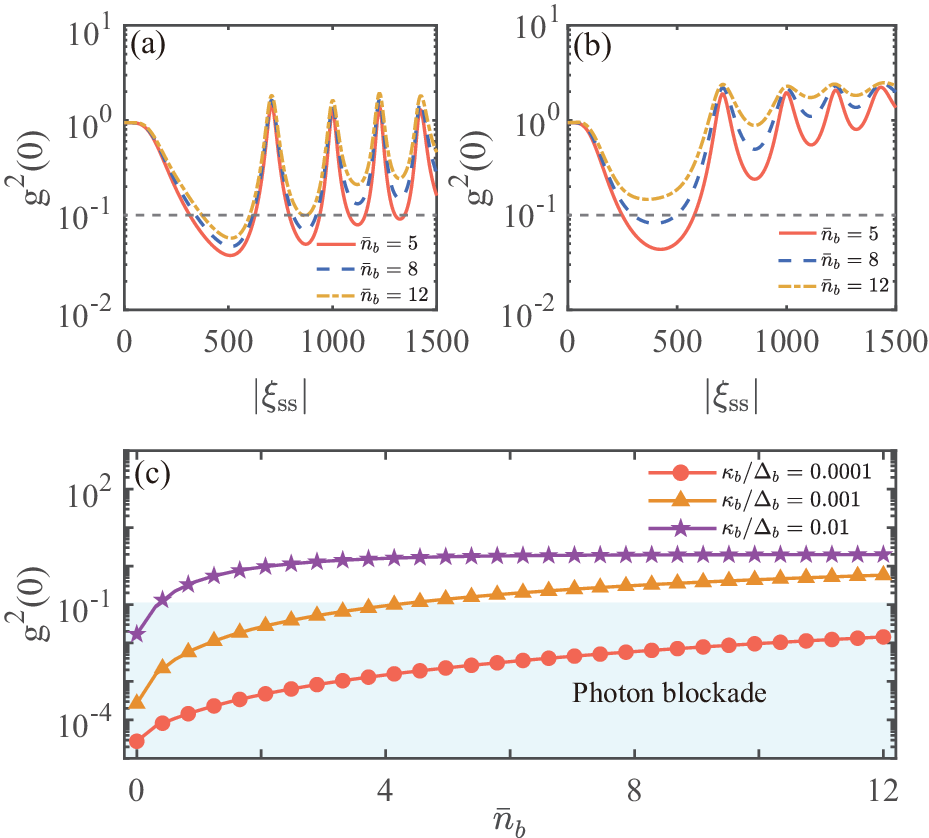}
\caption{(a) and (b) Plot of the equal-time second-order correlation function $g^{(2)}(0)$ as a function of $\vert\xi_{\text{ss}}\vert$ when $\bar{n}_{b}=5$, 8, and 12 at single-photon resonance $\Delta_{a}=g_{0}^2/\Delta_{b}$. Here the decay rates of mode $b$ are (a) $\kappa_{b}/\Delta_{b}=0.001$ and $\kappa_{b}/\Delta_{b}=0.01$. The other parameters are $\kappa_{a}/\Delta_{b}=0.03$ and $\kappa_{c}/\Delta_{b}=0.03$. (c) Plot of the correlation function $g^{(2)}(0)$ as a function of the thermal occupation number $\bar{n}_{b}$ when $\kappa_{b}$ takes different values. Here the driving detuning $\Delta_{a}=g_{0}^2/\Delta_{b}$ is chosen such that the single-photon transition is resonant. The other parameters are $\vert\xi_{\text{ss}}\vert=2000$ and $\kappa_{a}/\Delta_{b}=\kappa_{c}/\Delta_{b}=0.001$. In all panels, $g/\Delta_{b}=0.001$, $\Delta_{c}/\Delta_{b}=20$, $\bar{n}_{a}=\bar{n}_{c}=0$, and $\Omega_{a}/\kappa_{a}=0.1$.}
\label{g2vsnbfig}
\end{figure}
%%%%%%%%%%%%%%%%%%%%%%%%%%%%%

In Fig.~\ref{anaandnumg2fig}(a) we investigate the dependence of $g^{(2)}(0)$ as a function of the driving detuning $\Delta_{a}/\Delta_{b}$ to seek the optimal driving detuning. Here the solid curve is plotted based on the numerical solution of the quantum master equation, while the circles are plotted using the analytical solution. We find that the numerical results match well with the analytical one. As shown in Fig.~\ref{anaandnumg2fig}(a), the locations of these dips and peaks of $g^{(2)}(0)$ correspond to single- and two-photon resonant transitions, respectively, where the vertical dotted line is used to mark the single-photon resonance point $\Delta_{a}=g_{0}^{2}/\Delta_{b}$. This indicates that the single-photon transition $|0\rangle_{a}|0\rangle_{b}|0\rangle_{c}\leftrightarrow|1\rangle_{a}|\tilde{0}(1)\rangle_{b}|0\rangle_{c}$ (determined by $E_{1,0,0}\leftrightarrow E_{0,0,0}=0$) is resonant.
\begin{table*}[htb]
\centering
\caption{Parameters of the cross-Kerr-type coupled systems reported in literatures: the resonance frequencies $\omega_{a}$, $\omega_{b}$, and $\omega_{c}$ of modes $a$, $b$, and $c$, the cross-Kerr interaction strength $g$, the decay rates $\kappa_{a}$, $\kappa_{b}$, and $\kappa_{c}$ of modes $a$, $b$, and $c$, the thermal excitation occupations $\bar{n}_{a}$, $\bar{n}_{b}$, and $\bar{n}_{c}$ in the baths of modes $a$, $b$, and $c$. The reference in column 1 are experimental works.}
\label{table1sm}
\begin{tabular*}{1\textwidth}{@{\extracolsep{\fill}}c c c c c c c c c c c c}
\toprule
Ref. & Description &  $\frac{\omega_{a}}{2\pi}$ (GHz) & $\frac{\omega_{b}}{2\pi}$ (GHz) & $\frac{\omega_{c}}{2\pi}$ (GHz)
& $\frac{g}{2\pi}$ (kHz)  & $\frac{\kappa_{a}}{2\pi}$ (kHz)& $\frac{\kappa_{b}}{2\pi}$ (kHz) & $\frac{\kappa_{c}}{2\pi}$ (kHz)
& $\bar{n}_{a}$ & $\bar{n}_{b}$ & $\bar{n}_{c}$ \\
\hline
\cite{Gao2019Nature}$ $  & circuit QED & $4.944$ & $6.548$ & $5.476$ & $1.26\times 10^{3}$ & 10$-$15 &2$-$5&2$-$5& $\approx$0  & $\approx$0 & $\approx$0 \\
\hline
\cite{Schoelkopf2015}$ $ &  circuit QED & $8.493$ & $9.32$ & $7.249$ & $2.59\times10^{3}$ & $1.25$   & $5.25$  & $5.25$ & $\approx$0 & $\approx$0 & $\approx$0 \\
\hline
\botrule
\end{tabular*}
\end{table*}

To see the dependence of the photon blockade effect on the sideband resolution condition, in Fig.~\ref{anaandnumg2fig}(b) we illustrate $\log_{10}g^{(2)}(0)$ as a function of the parameters $\vert\xi_{\text{ss}}\vert$ and $\kappa_{a}$ for the single-photon resonance condition $\Delta_{a}=g_{0}^{(2)}/\Delta_{b}$~\cite{numericalcal1,numericalcal2}. Here we can see that $\log_{10}g^{(2)}(0)$ increases with the increase of the decay rate $\kappa_{a}$. When the decay rate $\kappa_{a}$ takes a small value, $\log_{10}g^{(2)}(0)$ exhibits some resonance peaks, which are induced by the phononlike-sideband resonant transitions. As shown in Fig.~\ref{anaandnumg2fig}(b), in some parameter regions (the valley region) the single-photon transition is resonant and the two-photon transition is far off-resonance. Then the photon blockade effect can be observed in mode $a$.

Moreover, the influence of the environment of mode $b$ on photon blockade effect can also be seen from $g^{(2)}(0)$. In Figs.~\ref{g2vsnbfig}(a) and~\ref{g2vsnbfig}(b), we display $g^{(2)}(0)$ versus $\vert\xi_{\text{ss}}\vert$ when the thermal occupation number takes different values $\bar{n}_{b}=5$, $8$, and $12$. This shows that the photon blockade effect becomes weaker for a larger value of $\bar{n}_{b}$. For the small decay rate $\kappa_{b}/\Delta_{b}=0.001$ in Fig.~\ref{g2vsnbfig}(a), however, the photon blockade effect still can be observed for a relatively large thermal occupation number $\bar{n}_{b}$. This phenomenon can also be seen from Fig.~\ref{g2vsnbfig}(c), where the correlation function $g^{(2)}(0)$ is plotted as a function of the thermal occupation number at selected values of the scaled decay rates $\kappa_{b}/\Delta_{b}=0.0001$, 0.001, and 0.01. Figure~\ref{g2vsnbfig}(c) shows that $g^{(2)}(0)$ increases with the increase of the thermal occupation number $\bar{n}_{b}$. In addition, the correlation function $g^{(2)}(0)$ is larger for a larger value of $\kappa_{b}$, which means that the dissipation of the system will weaken the photon blockade effect.
\begin{table*}[t]
\centering
\caption{Parameters used in the numerical simulations: the resonance frequency $\omega_{a}$ of mode $a$; the driving detunings (the effective frequencies in the displaced representation) $\Delta_{b}=\omega_{b}-\omega_{L}$ and $\Delta_{c}=\omega_{c}-\omega_{L}$ of modes $b$ and $c$, respectively; the Fredkin-type interaction strength $g$; the dimensionless displacement amplitude $|\xi_{\text{ss}}|$; the single-photon optomechanical-coupling strength $g_{0}=g|\xi_{\text{ss}}|$; the decay rates $\kappa_{a}$, $\kappa_{b}$, and $\kappa_{c}$ of modes $a$, $b$, and $c$, respectively; the thermal occupation numbers $\bar{n}_{a}$, $\bar{n}_{b}$, and $\bar{n}_{c}$ in the reservoirs of modes $a$, $b$, and $c$, respectively; and the single-photon strong-coupling parameter $g_{0}/\kappa_{a}$.}
\label{table2sm}
\begin{tabular*}{1\textwidth}{@{\extracolsep{\fill}}c c c c}
\toprule
Notation & Remarks & Scaled parameters  &Parameters  \\

\hline
$\Delta_{b}$ &  frequency scale  & $1$ & $2\pi\times 10$ MHz \\
\hline
$\Delta_{c}$ & effective frequency of mode $c$  & $\Delta_{c}/\Delta_{b}=1-20$ & $2\pi\times$ ($10-200$) MHz \\
\hline
$g$   & $g/\Delta_{b}\ll1$ for approximation &  $g/\Delta_{b}=0.001-0.01$ & $2\pi\times$ ($10-100$) kHz \\
\hline
$|\xi_{\text{ss}}|$ & $|\xi_{\text{ss}}|\gg1$ for coupling enhancement  &  & $1000-2000$ or $100-200$ \\
\hline
$g_{0}=g|\xi_{\text{ss}}|$ & single-photon optomechanical-coupling strength  &  $g_{0}/\Delta_{b}\approx1-2$ & $2\pi\times$ ($10-20$) MHz  \\
\hline
$\kappa_{a}$  & decay rate of mode $a$  & $\kappa_{a}/\Delta_{b}=0.01-0.1$   &  $2\pi\times$ ($100-1000$) kHz \\
\hline
$\kappa_{b}$  & decay rate of mode $b$  & $\kappa_{b}/\Delta_{b}=0.01-0.1$   &  $2\pi\times$ ($100-1000$) kHz \\
\hline
$\kappa_{c}$  & decay rate of mode $c$  & $\kappa_{c}/\Delta_{b}=0.01-0.1$   &  $2\pi\times$ ($100-1000$) kHz \\
\hline
$\bar{n}_{a}$  & thermal excitation number of mode $a$  &   &  $0$  \\
\hline
$\bar{n}_{b}$  & thermal excitation number of mode $b$   &   & $0-12$   \\
\hline
$\bar{n}_{c}$  & thermal excitation number of mode $c$   &   & $0$   \\
\hline
$g_{0}/\kappa_{a}$ & single-photon strong-coupling parameter  &  $0-200$  &    \\
\hline
$\vert\beta\vert_{\max }$ & $\vert\beta\vert_{\max }=2g_{0}/\Delta _{b}$ for $|\langle 0|\beta\rangle|\ll 1$   &  &  $4$  \\
\hline
\botrule
\end{tabular*}
\end{table*}

\section{Discussions on the experimental implementation}\label{Discussions}
The key element for experimental implementation of this scheme is to realize the Fredkin-type interaction. It has been suggested that the quantum Fredkin interaction can be constructed with two beam-splitter couplings involving modes $b$ and $c$, and a cross-Kerr interaction between modes $a$ and $b$~\cite{Milburn1989,Patel2016,Gao2019Nature}. To show this idea, we first introduce the beam-splitter transformation to a cross-Kerr interaction
\begin{eqnarray}
\label{unitary transformation of QFG}
U_{\text{tcK}} &=&B_{bc}(\theta)e^{-i\chi ta^{\dagger}ab^{\dagger}b}B_{bc}^{\dagger}(\theta)\nonumber\\
&=&\exp[-i\chi a^{\dagger }at(b^{\dagger }\cos \theta +c^{\dagger }\sin
\theta )(b\cos \theta +c\sin \theta )],  \nonumber\\
&&
\end{eqnarray}
where the unitary transformation operator for the beam splitter is defined as~\cite{Chuangbook}
\begin{equation}
B_{bc}(\theta) =\exp[\theta(b^{\dagger }c-bc^{\dagger})].
\end{equation}
The transformation of the beam splitter on the operators $b$ and $c$ can be obtained as
\begin{subequations}
\begin{align}
B_{bc}(\theta)bB_{bc}^{\dagger}(\theta)&=b\cos\theta+c\sin\theta,\\
B_{bc}(\theta)cB_{bc}^{\dagger}(\theta)&=-b\sin\theta+c\cos\theta .
\end{align}
\end{subequations}
For a 50:50 beam splitter, the mixing angle $\theta=\pi/4$; then Eq.~(\ref{unitary transformation of QFG}) becomes
\begin{eqnarray}
U_{\text{tcK}} &=&\exp \left[ -i\frac{\chi t}{2}a^{\dagger }a(b^{\dagger
}+c^{\dagger })(b+c)\right]   \nonumber \\
&=&\exp \left[ -i\frac{\chi t}{2}a^{\dagger }a(b^{\dagger }b+c^{\dagger }c)
\right] \exp \left[ -i\frac{\chi t}{2}a^{\dagger }a(b^{\dagger }c+c^{\dagger
}b)\right],  \nonumber \\
&&
\end{eqnarray}
where we have used the relation $\lbrack (b^{\dagger }b+c^{\dagger }c),(b^{\dagger }c+c^{\dagger }b)]=0$. To obtain a pure Fredkin interaction, we can design an inverse transformation $\exp \left[i\frac{\chi t}{2}a^{\dagger }a(b^{\dagger }b+c^{\dagger}c)\right]$ to eliminate the term $\exp \left[-i\frac{\chi t}{2}a^{\dagger }a(b^{\dagger }b+c^{\dagger }c)\right]$; then the unitary transformation can be reduced to
\begin{equation}
\label{up of QFG}
U^{\prime }=\exp \left[ -i\frac{\chi t}{2}a^{\dagger }a(b^{\dagger}c+c^{\dagger}b)\right].
\end{equation}
In terms of Eq.~(\ref{up of QFG}), we know that an effective Fredkin-type interaction can be obtained, which is described by the Hamiltonian
\begin{equation}
H_{F}=\frac{\chi}{2}a^{\dagger}a(b^{\dagger}c+c^{\dagger}b).
\end{equation}
On the other hand, the Hamiltonian corresponding the unitary operator
\begin{equation}
\label{u of QFG}
U=\exp\left[-i\frac{\chi t}{2}a^{\dagger}a(b^{\dagger}+c^{\dagger})(b+c)\right]
\end{equation}
reads
\begin{equation}
\label{Hamilotnian of QFG}
H=\frac{\chi }{2}a^{\dagger }a(b^{\dagger }+c^{\dagger })(b+c).
\end{equation}
In this case, if we apply a strong driving field to mode $c$ and use the Bogoliubov approximation
\begin{equation}
c^{\dagger}\rightarrow c^{\dagger}+\xi,\hspace{0.5cm} c\rightarrow c+\xi^{\ast},
\end{equation}
then Eq.~(\ref{Hamilotnian of QFG}) becomes
\begin{eqnarray}
\label{H prime}
H^{\prime } &=&\frac{\chi }{2}a^{\dagger }a\left[ (\xi ^{\ast }b+\xi
b^{\dagger })+(\xi ^{\ast }c+\xi c^{\dagger })\right.   \nonumber \\
&&\left. +(b^{\dagger }+c^{\dagger })(b+c)+|\xi |^{2}\right] .
\end{eqnarray}
Using the RWA under the corresponding parameter conditions, we can discard the four-mode coupling terms to obtain
\begin{equation}
\label{reduce H prime}
H^{\prime }=\frac{\chi }{2}a^{\dagger }a[(\xi ^{\ast }b+\xi b^{\dagger})+(\xi ^{\ast }c+\xi c^{\dagger })],
\end{equation}
which is equivalent to the two-mode-driven Fredkin-type interaction. Note that the terms of mode $b$ commutate with those of mode $c$; therefore, the physical applications proposed in this work can also be obtained with the second method.

From Eq.~(\ref{unitary transformation of QFG}) we see that the Fredkin-type interaction can be derived from two beam-splitter couplings involving modes $b$ and $c$ and a cross-Kerr interaction between modes $a$ and $b$. The cross-Kerr interaction can be realized in various platforms, such as cavity QED systems~\cite{Sinclair2007,Sinclair2008,Matsko2003,Kimble1998}, circuit QED systems~\cite{Hu2011,Nigg2012,Bourassa2012,Hoi2013,Schoelkopf2015,Majer2007}, and optomechanical systems~\cite{Thompson2008,Sankey2010,Karuza2013}. In particular, a quantum Fredkin gate has been realized by a three-dimensional circuit QED system~\cite{Gao2019Nature}, in which the Fredkin-type interaction strength can reach $g\sim10^{3}$ kHz, and the frequencies of these three bosic modes used are $\omega_{a}=2\pi\times4.944$ GHz, $\omega_{b}=2\pi\times6.548$ GHz, and $\omega_{c}=2\pi\times5.467$ GHz. Also, the decay rates $\kappa_{b}\approx\kappa_{c}\sim 2\pi\times(2-5)$ kHz and $\kappa_{a}\sim 2\pi\times(10-15)$ kHz have been used in Ref.~\cite{Gao2019Nature} (cf. Table~\ref{table1sm}). By applying a strong driving field to mode $c$ with frequency $\omega_{L}$ adjacent to $\omega_{c}$, the difference of the detunings $\Delta_{b}$ and $\Delta_{c}$ is $|\Delta_{b}-\Delta_{c}|\sim 10^{3}$ MHz, which is larger than the Fredkin interaction strength $g$. In addition, the driving amplitude $\Omega_{c}$ can be tuned so that $\Delta_{b}\sim g\vert\xi_{\text{ss}}\vert$ is accessible. Therefore, the parameter conditions for the approximate Hamiltonian~(\ref{appHamiltonian}) can be satisfied in circuit QED systems. As shown in Table~\ref{table2sm}, we used these experimentally accessible parameters $g=2\pi\times(10-100)$ kHz, $\Delta_{b}=2\pi\times 10$ MHz, $\Delta_{c}=2\pi\times$ (10$-$100) MHz, and $g_{0}= 2\pi\times$ (10$-$20) MHz in our simulations. Finally, we want to emphasize that the differences between this paper and Ref.~\cite{Liao2020} exist in the starting physical model, driving fields, and applications. Meanwhile, this scheme is universal and an optomechanical interaction involving optical and mechanical modes can be obtained based on a Fredkin-type interaction involving one optical mode and two mechanical modes. Note that the implementation of optomechanical-type interactions with two microwave fields was proposed in Refs.~\cite{Johansson2014,Johansson2015} and realized recently in experiments~\cite{Steele2020}.

\section{Conclusion}\label{Conclusions}
We have proposed a reliable scheme to implement a quantum simulation of ultrastrong optomechanics based on the Fredkin interaction. We have shown that the simulated optomechanical interaction can enter the single-photon strong-coupling and ultrastrong-coupling regimes, and hence single-photon optomechanical effects can be observed in this system. We have proved that the distinct macroscopic Schr\"{o}dinger cat states can be created, and that the transition from weak to strong quantum measurement can be demonstrated based on the tunable optomechanical coupling. Our proposal not only provides an inspiration for clarification of the quantum measurement puzzle, but also paves the way for the study of ultrastrong optomechanics in quantum simulators.

\begin{acknowledgments}
J.-Q.L. would like to thank Prof. Chang-Liang Ren for helpful discussions on the weak measurement. J.-Q.L. was supported in part by National Natural Science Foundation of China (Grants No. 12175061, No. 11822501, No. 11774087, and No. 11935006), Hunan Science and Technology Plan Project (Grant No. 2017XK2018), and the Science and Technology Innovation Program of Hunan Province (Grants No. 2020RC4047 and No. 2021RC4029).
\end{acknowledgments}


\begin{thebibliography}{99}
%review on optomechanics
\bibitem{Kippenberg2008rev} T. J. Kippenberg and K. J. Vahala, Cavity Optomechanics: Back-Action at the Mesoscale, Science \textbf{321}, 1172 (2008).
\bibitem{Aspelmeyer2012rev} M. Aspelmeyer, P. Meystre, and K. Schwab, Quantum optomechanics, Phys. Today \textbf{65} (7), 29 (2012).
\bibitem{Aspelmeyer2014} M. Aspelmeyer, T. J. Kippenberg, and F. Marquardt, Cavity optomechanics, Rev. Mod. Phys. \textbf{86}, 1391 (2014).

%optomechanical photon blcokade
\bibitem{Rabl2011} P. Rabl, Photon Blockade Effect in Optomechanical Systems, Phys. Rev. Lett. \textbf{107}, 063601 (2011).
\bibitem{Nunnenkamp2011} A. Nunnenkamp, K. B{\o}rkje, and S. M. Girvin, Single-Photon Optomechanics, Phys. Rev. Lett. \textbf{107}, 063602 (2011).
\bibitem{Liao2012} J.-Q. Liao, H. K. Cheung, and C. K. Law, Spectrum of single-photon emission and scattering in cavity optomechanics, Phys. Rev. A \textbf{85}, 025803 (2012).
\bibitem{Liao2013} J.-Q. Liao and C. K. Law, Correlated two-photon scattering in cavity optomechanics, Phys. Rev. A \textbf{87}, 043809 (2013).
\bibitem{Liao2013PRA}  J.-Q. Liao and F. Nori, Photon blockade in quadratically coupled optomechanical systems, Phys. Rev. A \textbf{88}, 023853 (2013).
\bibitem{Hong2013} T. Hong, H. Yang, H. Miao, and Y. Chen, Open quantum dynamics of single-photon optomechanical devices, Phys. Rev. A \textbf{88}, 023812 (2013).
\bibitem{Xu2013} X.-W. Xu, Y.-J. Li, and Y.-x. Liu, Photon-induced tunneling in optomechanical systems, Phys. Rev. A \textbf{87}, 025803 (2013).
\bibitem{Tang2014}  H. X. Tang and D. Vitali, Prospect of detecting single-photonforce effects in cavity optomechanics, Phys. Rev. A \textbf{89}, 063821 (2014).
%mechanical cat state generation
\bibitem{Marshall2003}  W. Marshall, C. Simon, R. Penrose, and D. Bouwmeester, Towards Quantum Superpositions of a Mirror, Phys. Rev. Lett. \textbf{91}, 130401 (2003).
\bibitem{Liao2016} J.-Q. Liao and L. Tian, Macroscopic Quantum Superposition in Cavity Optomechanics, Phys. Rev. Lett. \textbf{116}, 163602 (2016).

%%%%%quantum measurement
\bibitem{Neumann2018}  J. von Neumann,  \emph{Mathematical Foundations of Quantum Mechanics} (Princeton University Press, Princeton, 2018).

%enhencement of optomechanical coupling
\bibitem{Brennecke2008} F. Brennecke, S. Ritter, T. Donner, and T. Esslinger, Cavity Optomechanics with a
Bose-Einstein Condensate, Science \textbf{322}, 235 (2008).
\bibitem{Xuereb2012}  A. Xuereb, C. Genes, and A. Dantan, Strong Coupling and Long-Range Collective Interactions in Optomechanical Arrays, Phys. Rev. Lett. \textbf{109}, 223601 (2012).
\bibitem{Rimberg2014} A. J. Rimberg, M. P. Blencowe, A. D. Armour, and P. D. Nation, A cavity-Cooper pair transistor scheme for investigating quantum optomechanics in the ultra-strong coupling regime, New J. Phys. \textbf{16}, 055008 (2014).
\bibitem{Heikkila2014} T. T. Heikkil\"{a}, F. Massel, J. Tuorila, R. Khan, and M. A. Sillanp\"{a}\"{a}, Enhancing Optomechanical Coupling via the Josephson Effect, Phys. Rev. Lett. \textbf{112}, 203603 (2014).

\bibitem{Pirkkalainen2015} J.-M. Pirkkalainen, S. U. Cho, F. Massel, J. Tuorila, T. T. Heikkil\"{a}, P. J. Hakonen, and M. A. Sillanp\"{a}\"{a}, Cavity optomechanics mediated by a quantum two-level system, Nat. Commun. \textbf{6}, 6981 (2015).

\bibitem{Liao2014} J.-Q. Liao, K. Jacobs, F. Nori, and R. W. Simmonds, Modulated electromechanics: large enhancements of nonlinearities, New J. Phys. \textbf{16}, 072001 (2014).
\bibitem{Liao2015}  J.-Q. Liao, C. K. Law, L.-M. Kuang, and F. Nori, Enhancement of mechanical effects of single photons in modulated two-mode optomechanics, Phys. Rev. A \textbf{92}, 013822 (2015).
\bibitem{Lue2015}  X.-Y. L\"{u}, Y. Wu, J. R. Johansson, H. Jing, J. Zhang, and F. Nori, Squeezed Optomechanics with Phase-Matched Amplification and Dissipation, Phys. Rev. Lett. \textbf{114}, 093602 (2015).
\bibitem{Lemonde2016} M.-A. Lemonde, N. Didier, and A. A. Clerk, Enhanced nonlinear interactions in quantum optomechanics via mechanical amplification, Nat. Commun. \textbf{7}, 11338 (2016).
\bibitem{Li2016} P.-B. Li, H.-R. Li, and F.-L. Li, Enhanced electromechanical coupling of a nanomechanical resonator to coupled superconducting cavities, Sci. Rep. \textbf{6}, 19065 (2016).
\bibitem{Liao2020} J.-Q. Liao, J.-F. Huang, L. Tian, L.-M. Kuang, and C. P. Sun, Generalized ultrastrong optomechanical-like coupling, Phy. Rev. A \textbf{101}, 063802 (2020).
\bibitem{Wang2017} Z. Wang and A. H. Safavi-Naeini, Enhancing a slow and weak optomechanical nonlinearity with delayed quantum feedback, Nat. Commun. \textbf{8}, 15886 (2017).
\bibitem{Nori2014} I. M. Georgescu, S. Ashhab, and F. Nori, Quantum simulation, Rev. Mod. Phys. \textbf{86}, 153 (2014).
\bibitem{Hu2015}     D. Hu, S.-Y. Huang, J.-Q. Liao, L. Tian, and H.-S. Goan, Quantum coherence in ultrastrong optomechanics, Phys. Rev. A \textbf{91}, 013812 (2015).
%%%%%%weak measurement
\bibitem{Aharonov1998}    Y. Aharonov, D. Z. Albert, and L. Vaidman, How the result of a measurement of a component of the spin of a spin-$\frac{1}{2}$ particle can turn out to be 100, Phys. Rev. Lett. \textbf{60}, 1351 (1988).
\bibitem{Pan2020}    Y.-M. Pan, J. Zhang, E. Cohen, C.-W. Wu, P.-X. Chen, and N. Davidson, Weak-to-strong transition of quantum measurement in a trapped-ion system, Nat. Phys. \textbf{16}, 1206 (2020).

%%%A quantum Fredkin gate circuit
\bibitem{Milburn1989}  G. J. Milburn, Quantum optical Fredkin gate, Phys. Rev. Lett. \textbf{62}, 2124 (1989).
\bibitem{Patel2016}   R. B. Patel, J. Ho, F. Ferreyrol, T. C. Ralph, and G. J. Pryde, A quantum Fredkin gate, Sci. Adv. \textbf{2}, e1501531 (2016).
\bibitem{Gao2019Nature} Y. Y. Gao, B. J. Lester, K. S. Chou, L. Frunzio, M. H. Devoret, L. Jiang, S. M. Girvin, and R. J. Schoelkopf, Entanglement of bosonic modes through an engineered exchange interaction, Nature (London) \textbf{566}, 509 (2019).
\bibitem{Ludwig2008} M. Ludwig, B. Kubala, and F. Marquardt, The optomechanical instability in the quantum regime, New J. Phys. \textbf{10}, 095013 (2008).

%%%F-C factor
\bibitem{Franck1925} J. Franck, Elementary processes of photochemical reactions, Trans. Faraday Soc. \textbf{21}, 536 (1925).
\bibitem{Condon1926} E. Condon, A Theory of Intensity Distribution in Band Systems, Phys. Rev. \textbf{28}, 1182 (1926).
\bibitem{Leturcq2009}  R. Leturcq, C. Stampfer, K. Inderbitzin, L. Durrer, C. Hierold, E. Mariani, M. G. Schultz, F. von Oppen, and K. Ensslin, Franck-Condon blockade in suspended carbon nanotube quantum dots, Nat. Phys. \textbf{5}, 327 (2009).

%%%%Book
\bibitem{Barnettbook}   S. M. Barnett and P. M. Radmore, \emph{Methods in Theoretical Quantum Optics} (Clarendon,  Oxford, 1997).
\bibitem{Milburnbook}  D. F. Walls and G. J. Milburn, \emph{Quantum Optics} (Springer, Berlin, 2008).
%%%%%weak measurement in optomechanics
\bibitem{Pepper2012} B. Pepper, R. Ghobadi, E. Jeffrey, C. Simon, and D. Bouwmeester, Optomechanical Superpositions via Nested Interferometry, Phys. Rev. Lett. \textbf{109}, 023601 (2012).
\bibitem{Kofman2012}  A. G. Kofman, S. Ashhab, and F. Nori, Nonperturbative theory of weak pre-and post-selected measurements, Phys. Rep. \textbf{520}, 43 (2012).
\bibitem{Dressel2014} J. Dressel, M. Malik, F. M. Miatto, A. N. Jordan, and R. W. Boyd, Colloquium: Understanding quantum weak values: Basics and applications, Rev. Mod. Phys. \textbf{86}, 307 (2014).

%%%%%%%%
\bibitem{numericalcal1}   J. R. Johansson, P. D. Nation, and F. Nori, QuTiP: An opensource Python framework for the dynamics of open
quantum systems, Comput. Phys. Commun. \textbf{183}, 1760 (2012).
\bibitem{numericalcal2}  J. R. Johansson, P. D. Nation, and F. Nori, QuTiP 2: A Python framework for the dynamics of open quantum systems, Comput. Phys. Commun. \textbf{184}, 1234 (2013).

\bibitem{Chuangbook}  M. A. Nielsen and I. L. Chuang, \emph{Quantum Computation and Quantum Information} (Cambridge University Press, Cambridge, 2000).
%atoms coupled to two fields
\bibitem{Sinclair2007} G. F. Sinclair and N. Korolkova, Cross-Kerr interaction in a four-level atomic system, Phys. Rev. A \textbf{76}, 033803 (2007).
\bibitem{Sinclair2008} G. F. Sinclair and N. Korolkova, Effective cross-Kerr Hamiltonian for a nonresonant four-level atom, Phys. Rev. A \textbf{77}, 033843 (2008).

\bibitem{Matsko2003}     A. B. Matsko, I. Novikova, G. R. Welch, and M. S. Zubairy, Enhancement of Kerr nonlinearity by multiphoton coherence, Opt. Lett. \textbf{28}, 96 (2003).

\bibitem{Kimble1998}       H. J. Kimble, Strong Interactions of Single Atoms and Photons in Cavity QED, Phys. Scr.  \textbf{1998}, 127 (1998).

% circuit QED system
\bibitem{Hu2011}            Y. Hu, G.-Q. Ge, S. Chen, X.-F. Yang, and Y.-L. Chen, Cross-Kerr-effect induced by coupled Josephson qubits in circuit quantum electrodynamics, Phys. Rev. A \textbf{84}, 012329 (2011).

\bibitem{Nigg2012}          S. E. Nigg, H. Paik, B. Vlastakis, G. Kirchmair, S. Shankar, L. Frunzio, M. H. Devoret, R. J. Schoelkopf, and S. M. Girvin, Black-Box Superconducting Circuit Quantization, Phys. Rev. Lett. \textbf{108}, 240502 (2012).

\bibitem{Bourassa2012}     J. Bourassa, F. Beaudoin, J. M. Gambetta, and A. Blais, Josephson-junction-embedded transmission-line resonators: From Kerr medium to in-line transmon, Phys. Rev. A \textbf{86}, 013814 (2012).

\bibitem{Hoi2013}   I. C. Hoi, A. F. Kockum, T. Palomaki, T. M. Stace, B. Fan, L. Tornberg, S. R. Sathyamoorthy, G. Johansson, P. Delsing, and C. M. Wilson, Giant Cross-Kerr Effect for Propagating Microwaves Induced by an Artificial Atom, Phys. Rev. Lett. \textbf{111}, 053601 (2013).

\bibitem{Schoelkopf2015}     E. T. Holland, B. Vlastakis, R. W. Heeres, M. J. Reagor, U. Vool, Z. Leghtas, L. Frunzio, G. Kirchmair, M. H. Devoret, M. Mirrahimi, and R. J. Schoelkopf, Single-Photon-Resolved Cross-Kerr Interaction for Autonomous Stabilization
of Photon-Number States, Phys. Rev. Lett. \textbf{115}, 180501 (2015).

\bibitem{Majer2007}     J. Majer, J. M. Chow, J. M. Gambetta, Jens Koch, B. R. Johnson, J. A. Schreier, L. Frunzio, D. I. Schuster, A. A. Houck, A. Wallraff, A. Blais, M. H. Devoret, S. M. Girvin, and R. J. Schoelkopf, Coupling superconducting qubits via a cavity bus, Nature (London) \textbf{449}, 443  (2007).
%membrane-in-the-middle configuration
\bibitem{Thompson2008}     J. D. Thompson, B. M. Zwickl, A. M. Jayich, F. Marquardt, S. M. Girvin, and J. G. E. Harris, Strong dispersive coupling of a high-finesse cavity to a micromechanical membrane, Nature (London) \textbf{452}, 72 (2008).
\bibitem{Sankey2010}       J. C. Sankey, C. Yang, B. M. Zwickl, A. M. Jayich, and J. G. E. Harris, Strong and tunable nonlinear optomechanical coupling in a low-loss system, Nat. Phys. \textbf{6}, 707 (2010).
\bibitem{Karuza2013}       M. Karuza, M. Galassi, C. Biancofiore, C. Molinelli, R. Natali, P. Tombesi, G. Di Giuseppe, and D. Vitali, Tunable linear and quadratic optomechanical coupling for a tilted membrane within an optical cavity: theory and experiment, J. Opt. \textbf{15}, 025704 (2013).
\bibitem{Johansson2014} J. R. Johansson, G. Johansson, and F. Nori, Optomechanical-like coupling between superconducting resonators, Phys. Rev. A \textbf{90}, 053833 (2014).
\bibitem{Johansson2015} E.-j. Kim, J. R. Johansson, and F. Nori, Circuit analog of quadratic optomechanics, Phys. Rev. A \textbf{91}, 033835 (2015).
\bibitem{Steele2020}     D. Bothner, I. C. Rodrigues, and G. A. Steele, Photon-pressure strong coupling between two superconducting circuits,   Nat. Phys. \textbf{17}, 85 (2020).
\end{thebibliography}
\end{document}